\documentclass[prc,floatfix,groupedaddress,nofootinbib,showpacs,preprintnumbers,
amsmath,amssymb,amsfonts,superscriptaddress,widetable] {revtex4-1}
\usepackage{graphicx}
\usepackage{dcolumn}
\usepackage{mathrsfs,amsmath,amssymb}
\usepackage{bm}
\usepackage{graphicx}
\usepackage{dcolumn}
\usepackage{bm}
\usepackage{array}
\usepackage{bigstrut}
\usepackage[usenames]{color}
\begin{document}

\title{Nuclear charge radii: Density functional theory  \\
 meets Bayesian neural networks}
\author{R. Utama}
\email{ru11@my.fsu.edu} 
\affiliation{Department of Physics, Florida State University, Tallahassee, FL 32306} 
\author{Wei-Chia Chen}
\email{wc09c@my.fsu.edu} 
\affiliation{Department of Physics, Florida State University, Tallahassee, FL 32306}
\author{J. Piekarewicz}
\email{jpiekarewicz@fsu.edu}
\affiliation{Department of Physics, Florida State University, Tallahassee, FL 32306}
\date{\today}
\begin{abstract}
\begin{description}
\item[Background]  The distribution of electric charge in atomic nuclei is fundamental 
                               to our understanding of the complex nuclear dynamics and a
                               quintessential observable to validate nuclear structure models.      
\item[Purpose]        To explore a novel approach that combines sophisticated
	                       models of nuclear structure with Bayesian neural networks 
	                       (BNN) to generate predictions for the charge radii of 
	                       thousands of nuclei throughout the nuclear chart.
\item[Methods]        A class of relativistic energy density functionals is used
	                       to provide robust predictions for nuclear charge radii. In
	                       turn, these predictions are refined through 
	                       Bayesian learning for a neural network that is trained 
	                       using residuals between theoretical predictions and 
	                       the experimental data.
\item[Results]         Although predictions obtained with density functional
	                      theory provide a fairly good description of experiment,
	                      our results show significant improvement (better than
	                      40\%) after BNN refinement. Moreover, these improved 
	                      results for nuclear charge radii are supplemented with 
	                      theoretical error bars. 
\item[Conclusions] We have successfully demonstrated the ability of the BNN 
                               approach to significantly increase the accuracy of nuclear 
                               models in the predictions of nuclear charge radii. However,
                               as many before us, we failed to uncover the underlying
                               physics behind the intriguing behavior of charge radii 
                               along the calcium isotopic chain.
\end{description}
\end{abstract}
\pacs{21.60.Jz,24.10.Jv,21.10.Ft} 

\maketitle

\section{Introduction}
\label{intro}

Nuclear saturation, the existence of an equilibrium density, is one of
the most fundamental manifestations of the complex nuclear
dynamics. The successful liquid drop model (LDM), formulated by Bethe
and Weizs\"acker shortly after the discovery of the neutron by
Chadwick, is firmly rooted in the existence of an equilibrium
density\,\cite{Weizsacker:1935,Bethe:1936}.  In the LDM, the nucleus is
treated as an incompressible quantum drop of uniform density
$\rho_{0}\!\simeq\!0.15\,{\rm fm}^{-3}$ consisting of $Z$ protons, $N$
neutrons, and a total baryon number of $A\!=\!N\!+\!Z$.  Among the
best-known consequences of nuclear saturation is the fact that the
average size of the nucleus scales with the total number of nucleons
as $A^{1/3}$. That is,
\begin{equation}
 R_{0}(A)\!=r_{{}_{\!0}}A^{1/3} \;\;{\rm where}\;
 r_{{}_{\!0}}\!=\!\left(\frac{3}{4\pi\rho_{0}}\right)^{\!1/3}\!\!\!\simeq\!1.15\,{\rm fm}\,.
 \label{Radius}
\end{equation}
In turn, the root-mean-square (RMS) charge radius of such a uniform 
distribution is given by
\begin{equation}
 R_{\rm ch}^{\rm LDM}(A)=\sqrt{\frac{3}{5}}\,r_{{}_{\!0}}A^{1/3}\,.
 \label{RMSRadius}
\end{equation}

In an effort to improve the quantitative standing of the LDM,
microscopic-macroscopic (``mic-mac'') models have been developed to
account for the physics that is missing from such a simple
description. Although most of these efforts have been devoted to
improve nuclear-mass
predictions\,\cite{Moller:1981zz,Moller:1988,Moller:1993ed,Duflo:1995},
significant activity has also been dedicated to refine the description
of charge radii.  For stable heavy nuclei with a large volume-to-surface 
ratio, the LDM estimates provide a qualitative---and often
quantitative---description of experimental charge radii. However,
given its description of the nucleus as a uniform liquid drop with a
sharp density, the LDM fails to reproduce the charge radii of light
nuclei that are dominated by surface effects\,\cite{Brown:1984}. In an
attempt to account for the nuclear surface, Myers and Schmidt
introduced a gaussian modification to the sharp density of the form
$\exp({-r^2/2b^2})$, with a best-fit parameter $b\!=\!0.99\,{\rm fm}$ independent of $N$
and $Z$\,\cite{Myers:1983}. Such form was
inspired by the Helm form factor that was introduced six decades ago
to analyze elastic scattering of electrons from
nuclei\,\cite{Helm:1956zz}. The Helm form factor, namely, the Fourier
transform of the underlying charge density, is defined as the product
of two fairly simple form factors: one associated with a uniform sharp
density and the other one with a gaussian distribution. In the refined 
model of Myers and Schmidt, referred hereafter as the 
``extended-liquid-drop'' (ELD) model\,\cite{Brown:1984}, the RMS 
charge radius of a nucleus with mass number $A$ is given by
\begin{equation}
 R_{\rm ch}^{\rm {ELD}}(A) = \sqrt{\frac{3}{5}}\,r_{{}_{\!0}}A^{1/3}
 \left(1+\alpha A^{-2/3}-\beta A^{-4/3}\right),
\label{ELD}  
\end{equation} 
where $\alpha\!=\!1.57$ and $\beta\!=\!1.04$ were obtained by fitting
to the then available experimental data\,\cite{Myers:1983}. Although
the ELD provides a definite improvement over the original LDM, its
sole dependence on $A$, rather than on both $Z$ and $A$, suggests 
that some serious deficiencies remain. Nevertheless, the ELD model
provides a useful baseline to test our approach. As a more recent 
mic-mac representative, we include the simple, yet fairly successful, 
model by Zhang and collaborators\,\cite{Zhang:2002}. In this model 
the charge radius is parametrized as follows:
\begin{equation}
 R_{\rm ch}(Z,A) = r_{{}_{\!A}}A^{1/3}
 \left[1-b\left(\frac{N\!-\!Z}{A}\right)+\frac{c}{A}\right],
\label{Zhang4}  
\end{equation} 
where $r_{{}_{\!A}}\!=\!0.966\,{\rm fm}$, $b\!=\!0.182$ and 
$c\!=\!1.652$. Note that whereas the adopted model is from
Ref.\,\cite{Zhang:2002}, the model parameters are from the 
more recent---and accurate---calibration by Bayram {\sl et al.}; 
see Table I of Ref.\,\cite{Bayram:2013rad}.

However, we expect that our suggested approach will find its best expression 
in a method that combines predictions from an accurately-calibrated density
functional which are subsequently refined through the use of a Bayesian Neural 
Network (BNN). Although enormous progress has been made in the design of
nuclear energy density functionals\,\cite{FAYANS:2000,Goriely:2010bm,
Kortelainen:2010hv,Erler:2012qd,Chen:2014sca,Chen:2014mza}, the reality 
is that some lingering 
discrepancy between theory and experiment is inevitable. The aim of Bayesian 
learning is to eliminate, or at least to significantly reduce, these discrepancies. 
Such an approach has already been successfully tested in the case of nuclear 
masses of relevance to the composition of the neutron-star 
crust\,\cite{Utama:2015hva}. Motivated by some intriguing
new measurements in the calcium isotopes\,\cite{Ruiz:2016gne}, we
extend our formalism to predict nuclear charge radii across the whole
nuclear landscape.  In essence, our underlying theoretical approach is
rooted in Strutinsky's energy theorem\,\cite{Brack:1997} which states
that the nuclear binding energy may be separated into two components:
one large and smooth and the other one small and
fluctuating\,\cite{Strutinsky:1967}. Moreover, we successfully
conjectured that Strutinsky's energy theorem may be extended to any
nuclear observable that also displays slowly varying
dynamics\,\cite{Piekarewicz:2009av}. In particular, it was shown that
the \emph{Garvey-Kelson relations}\,\cite{Garvey:1966zz,GARVEY:1969zz,
Preston:1993} provide a fruitful framework for the prediction of
nuclear charge radii\,\cite{Piekarewicz:2009av}. Indeed, besides
successfully testing the Garvey-Kelson relations (GKRs) on hundreds of
nuclei whose charge radius is experimentally known, we made
predictions for 116 nuclei whose charge radius was unknown at
that time\,\cite{Piekarewicz:2009av}; for a similar treatment see
Ref.\,\cite{Sun:2014}.  Unfortunately, the GKRs are \emph{local}
relations whose success depends critically on the knowledge of the
landscape around the nucleus of interest. Hence, though enormously
successful when measurements on the nearest neighbors to the nucleus
of interest exist, the GKRs extrapolate poorly. Thus, it is our hope
that by properly implementing a \emph{global} description through
Bayesian learning of neural networks some of these deficiencies will
be mitigated.

Together with nuclear masses, nuclear sizes are among the most
fundamental properties of atomic nuclei. With the commissioning of
radioactive beam facilities all over the world, understanding the
limits of nuclear existence, particularly the evolution of nuclear
masses and sizes, represents one of the overarching questions that
animate nuclear physics today.  Regarding the charge radii of nuclei
with a large neutron excess, a recent laser spectroscopy experiment at
ISOLDE, CERN has observed ``unexpectedly large charge radii of
neutron-rich calcium isotopes''\,\cite{Ruiz:2016gne}. Although
considerable theoretical progress has been made in reproducing---and
predicting---the evolution of nuclear masses along the calcium
isotopes\,\cite{Holt:2010yb,Hagen:2012fb,Chen:2014mza}, predicting the
complex evolution of the corresponding charge radii remains a serious
challenge for theoretical descriptions that range from \emph{ab
initio} methods to density functional theory (DFT)\,\cite{Ruiz:2016gne}. In an
effort to gain some insights into this problem, we extend the (BNN)
approach implemented for the first time in
Ref.\,\cite{Piekarewicz:2009av} to nuclear charge radii. In doing so,
we follow closely the ideas outlined in the text by R.M. Neal entitled
\emph{Bayesian Learning for Neural Networks}\,\cite{Neal1996}. To
explore the foundations and applications of artificial neural networks
see Refs.\,\cite{Bishop1995,Haykin1999,Vapnik1998}. We note that the
first application of artificial neural networks to nuclear physics
dates back to the early 1990s and continues to this day in the work of
Clark and collaborators\,\cite{Gazula:1992,Gernoth:1993,Gernoth:1995,
Clark:1999,Athanassopoulos:2003qe,Athanassopoulos:2005rc,
Clark:2006ua,Costiris:2009}. Artificial neural networks have also been used
in these studies to reproduce, among other things, (a) the \emph{differences} 
between experimental nuclear masses and theoretical predictions provided 
by the Finite Range Droplet Models (FRDM) and (b) $\beta$-decay rates 
of relevance to r-process nucleosynthesis. More recently, artificial 
neural networks have been used in the study of binding-energy 
systematics\,\cite{Bayram:2013hi} and nuclear charge 
radii\,\cite{Akkoyun:2012yf}.

 We have organized the paper as follows. In Sec.\,\ref{Formalism} we
describe briefly the relativistic density functional that is used to
predict charge radii throughout the nuclear chart.  Following this
discussion, we review the main ideas involved in training and
validating a neural network by concentrating on the \emph{residuals}
between theoretical predictions and experimental measurements of
charge radii, as obtained from the latest compilations by Angeli and
collaborators\,\cite{Angeli:2004,Angeli:2013}. We then proceed to
Sec.\,\ref{Results} to display the results of our calculations with a
special emphasis on the BNN refinement of the ``bare'' model
predictions. Finally, in Sec.\,\ref{Conclusions} we summarize our work.


\section{Formalism}
\label{Formalism}

The formalism section is composed of two subsections that briefly
address topics that have been discussed in great detail in the
existent literature. These are: (a) the calibration and predictions of
a particular class of relativistic density functionals and (b) the
Bayesian learning approach for neural networks.

\subsection{Relativistic Energy Density Functionals}
\label{REDF}

The effective relativistic model employed here is inspired in the
original work of Walecka from the mid 1970s\,\cite{Walecka:1974qa} and
continues to this day with refinements that systematically improve the
quantitative predictions of the model. Besides providing an accurate
description of finite nuclei, the relativistic approach offers a
Lorentz covariant extrapolation to high-density matter, a critical
fact in the simulation of neutron stars. In such a framework, the
basic degrees of freedom include nucleons that interact via the
exchange of three ``mesons'' and the photon. The effective
Lagrangian density for this class of models is given
by\,\cite{Walecka:1974qa,Boguta:1977xi,Horowitz:1981xw,
Serot:1984ey,Mueller:1996pm,Serot:1997xg,Lalazissis:1996rd,
Lalazissis:1999,Horowitz:2000xj,Todd-Rutel:2005fa,Chen:2014sca,
Chen:2014mza}:
\begin{eqnarray}
{\mathscr L}_{\rm int} &=&
\bar\psi \left[g_{\rm s}\phi   \!-\! 
         \left(g_{\rm v}V_\mu  \!+\!
    \frac{g_{\rho}}{2}{\mbox{\boldmath $\tau$}}\cdot{\bf b}_{\mu} 
                               \!+\!    
    \frac{e}{2}(1\!+\!\tau_{3})A_{\mu}\right)\gamma^{\mu}
         \right]\psi \nonumber \\
                   &-& 
    \frac{\kappa}{3!} (g_{\rm s}\phi)^3 \!-\!
    \frac{\lambda}{4!}(g_{\rm s}\phi)^4 \!+\!
    \frac{\zeta}{4!}   g_{\rm v}^4(V_{\mu}V^\mu)^2 +
   \Lambda_{\rm v}\Big(g_{\rho}^{2}\,{\bf b}_{\mu}\cdot{\bf b}^{\mu}\Big)
                           \Big(g_{\rm v}^{2}V_{\nu}V^{\nu}\Big)\;.
 \label{LDensity}
\end{eqnarray}
Note that here $\psi$ represents the isodoublet nucleon field,
$A_{\mu}$ the photon field, and $\phi$, $V_{\mu}$, and ${\bf b}_{\mu}$
the isoscalar-scalar $\sigma$-, isoscalar-vector $\omega$-, and
isovector-vector $\rho$-meson fields, respectively.  The first line of
the above equation contains the conventional Yukawa couplings between
the nucleons and the mesons.  In the original Walecka model
(also knows as the ``$\sigma$-$\omega$ model'') only the two isoscalar
mesons were considered\,\cite{Walecka:1974qa}. Nevertheless, such a
simple description was already able to provide a natural and
compelling explanation for the mechanism behind nuclear-matter
saturation. Later on, Horowitz and Serot added the isovector
$\rho$-meson and the photon to provide a highly improved description
of the ground-state properties of finite
nuclei\,\cite{Horowitz:1981xw}. The second line in
Eq.\,(\ref{LDensity}) includes nonlinear self and mixed interactions
among the meson fields that effectively generate many-body (or density-dependent)
 interactions.  The need for such kind of terms was
recognized early on by Boguta and Bodmer who introduced the two
isoscalar parameters, $\kappa$ and $\lambda$, to soften the equation
of state of symmetric nuclear matter in the vicinity of the saturation
density\,\cite{Boguta:1977xi}. In turn, $\zeta$ may be used to
calibrate the equation of state at high density from the astrophysical
determination of neutron-star masses. Note that this may be done
without sacrificing the good agreement with physical observables
sensitive to the equation of state near saturation
density\,\cite{Mueller:1996pm}. Finally, $\Lambda_{\rm v}$ is the only
parameter that involves a mixing between the isoscalar and isovector
sectors.  As such, $\Lambda_{\rm v}$ may be efficiently tuned to
soften the density dependence of symmetry energy, which is
traditionally stiff in relativistic mean-field models that contain
only one single isovector parameter.

\subsection{Bayesian Neural Networks}
\label{BNN} 

Despite the steady and considerable progress that has been achieved in
the design of nuclear mass models and energy density functionals,
often the accuracy achieved is inadequate to properly describe
astrophysical phenomena. For example, it appears that in order to
resolve the abundance pattern in r-process nucleosynthesis, mass
uncertainties must be reduced by factors of 3-5 from their current
limit (to $\lesssim100\,{\rm keV}$); see Ref.\,\cite{Mumpower:2015ova}
and references contained therein. Moreover, given the suggestion of an
inherent limit to the accuracy of mass
models\,\cite{Aberg:2012,Barea:2005fz}, a novel approach is sorely
needed.  The novel approach that we advocate here is Bayesian learning
for neural networks\,\cite{Neal1996}. In the particular case of
nuclear observables, we propose a combined scheme that relies on
accurate model predictions which are then refined by training a
suitable neural network on the \emph{residuals} between the
experimental data and the theoretical predictions. We have
successfully tested such paradigm in the case of nuclear
masses\,\cite{Utama:2015hva}.  In this work we extend the 
approach to the description of nuclear charge radii.

Although a detailed description of Bayesian neural networks goes
beyond the scope of this paper, we nevertheless highlight some of the
main features that are critical to the implementation. First,
multilayer feed-forward neural networks have been shown to be
``universal approximators", as they are capable of approximating
\emph{any} measurable function from one finite dimensional space to
another to any desired degree of accuracy\,\cite{Hornik:1989}. Given
that we advocate an approach where the first step is the design of a 
model that incorporates as much physics as possible, we expect that the
experiment-theory residuals will be a smooth function which may be
faithfully emulated with a relatively simple neural network. Second,
Bayesian learning requires specifying a \emph{prior distribution}
that captures our beliefs \emph{prior} to unveiling the data. Once the
data is unveiled, the \emph{posterior distribution} is used to make
predictions with properly quantified uncertainties about future
measurements. Note that the posterior distribution encodes the
improvement in our prior knowledge as a result of the new
data. Finally, the Bayesian approach with the selection of a robust
prior is less prone to overfitting the training
data\,\cite{Neal1996,Titterington:2004}; unfortunately, the prior
selection of neural network parameters has no obvious connection to
our prior physics knowledge\,\cite{Hornik:1989}.

Statistical inference based on Bayes' theorem connects a given
hypothesis (in terms of our beliefs for a set of parameters $\omega$)
and a set of data $(x,t)$ to a posterior probability $p(\omega|x,t)$
that is used to make predictions on future
data\,\cite{Stone:2013}. That is,
\begin{equation}
 p(\omega|x,t)=\frac{p(x,t|\omega)p(\omega)}{p(x,t)},
 \label{BayesRule}
\end{equation} 
where $p(x,t|\omega)$ is the ``likelihood" that a given model
describes the data and $p(\omega)$ is the prior distribution of the
parameters $\omega$.  Following common practices, we assume a Gaussian
distribution for the likelihood in terms of an objective (or ``cost'')
function obtained from a least-squares fit to the empirical data. That
is,
\begin{equation}
 p(x,t|\omega)=\exp\big(\!-\chi^{2}(\omega)/2\big),
 \label{likelihood}
\end{equation} 
where $\chi^{2}(\omega)$ is given by
\begin{equation}
 \chi^2(\omega)=\sum_{i=1}^{N}
 \left(\frac{t_i-f(x_i,\omega)}{\Delta t_{i}}\right)^2.
 \label{Chi2}
\end{equation}    
Here $N$ is the number of empirical data, $t_{i}\!\equiv\!t(x_{i})$ is
the {\sl i}th observable with $\Delta t_{i}$ its associated error, and
the function $f(x,\omega)$ (given below) depends on the input data $x$
and the model parameters $\omega$. In our particular case,
$x\!\equiv\!(Z,A)$ denotes the two input variables (proton and mass
numbers) and $t(x)\!\equiv\!\delta R_{\rm ch}(Z,A)$ the charge-radius
residual, namely, the difference between the experimental data and the
theoretical predictions as provided, for example, by a nuclear energy
density functional. Unlike other approaches, Bayesian predictions are
based on a large number of estimates of the model parameters that are
generated from the posterior distribution.  That is,
\begin{equation}
 \langle f_{n}\rangle = \int f(x_{n},\omega)p(\omega|x,t)\,d\omega
 =\frac{1}{K}\sum_{k=1}^{K} f(x_{n},\omega _{k}),
 \label{Avgfn}
\end{equation} 
where $x_{n}\!=\!(Z_{n},A_{n})$ represents a nucleus with charge
$Z_{n}$ and mass number $A_{n}$, and $f(x_{n},\omega)$ is the neural
network prediction for $\delta R_{\rm ch}(Z_{n},A_{n})$ for a given
set of parameters $\omega$. The Bayesian estimate of the average of
$\delta R_{\rm ch}(Z_{n},A_{n})$ is obtained by integrating over the
posterior parameter distribution using Markov Chain Monte Carlo (MCMC)
sampling\,\cite{Neal1996}; in Eq.\,(\ref{Avgfn}) $K$ refers to the
total number of Monte Carlo samples.  A distinct advantage of the
Bayesian method is that predictions for the averages can be
complemented with a proper quantification of the uncertainty. Indeed,
the statistical uncertainty is given by
\begin{equation}
 \Delta f_{n} = \sqrt{\langle f_{n}^{2}\rangle - \langle f_{n}\rangle^{2}},
 \label{Errorfn}
\end{equation} 
where $\langle f_{n}^{2}\rangle$ is evaluated following the same 
procedure described in Eq.\,(\ref{Avgfn}).

\begin{figure}[ht]
\vspace{-0.05in}
\includegraphics[width=0.45\columnwidth] {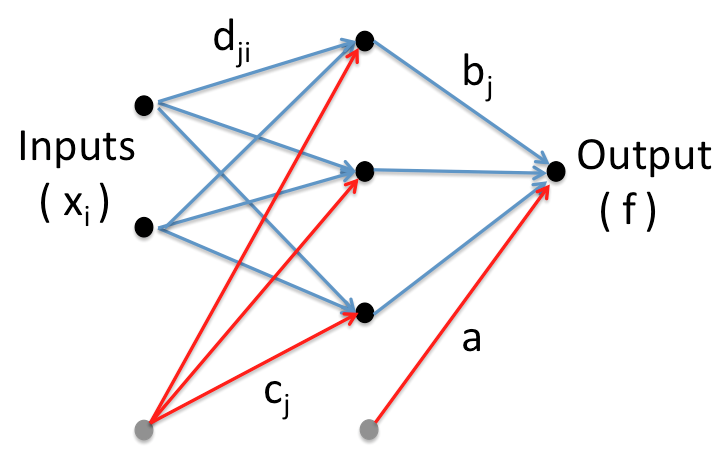}
\caption{An example of a feed-forward neural network with a single 
hidden layer consisting of three nodes. In our case, two inputs
define the nucleus of interest ($Z$ and $A$) and a single output
provides an estimate of $\delta R_{\rm ch}(Z,A)$.}
\label{Fig1}
\end{figure}

All that remains is to specify the form of the neural network function
(or ``emulator'') $f(x,\omega)$ and the prior distribution $p(\omega)$. 
Note that we can ignore the marginal likelihood $p(x,t)$ in 
Eq.\,(\ref{BayesRule}) since as far as the MCMC is concerned, it 
represents an overall normalization factor
independent of $\omega$.  In this work, as in our previous
one\,\cite{Utama:2015hva}, we use a feed-forward neural network 
model of the following form:
\begin{equation}
  f(x,\omega)=a+\sum_{j=1}^H b_j {\rm tanh}\left(c_j+\sum_{i=1}^I d_{ji} x_i\right),
  \label{ANN}
\end{equation}
where the model parameters are given by
$\omega\!=\!\big\{a,b_{j},c_{j},d_{ji}\big\}$, $H$ is the number of
hidden nodes, and $I$ is the number of inputs. For two input variables
(as in our case) the function in Eq.\,(\ref{ANN}) contains a total of
$1\!+\!4H$ parameters. Naturally, a one-size-fits-all prescription for
selecting the optimal number of hidden nodes $H$ is not available so a
considerable amount of trial and error is required.  An example of a
neural network consisting of a single hidden layer with three
nodes is shown in Fig.\,\ref{Fig1}.

Prior probabilities encode our beliefs concerning the model parameters
and are an essential ingredient of the Bayesian paradigm. This feature
is highly desirable as it allows us to inject our own physics biases
and intuition which are often well informed by previous data or
experiences.  Unfortunately, physics intuition is of no help in the
design of the connection weights ({\sl i.e.} model parameters) $\omega$. 
In this case, one must rely
on tried-and-true methods\,\cite{Neal1996}. Following our earlier
work\,\cite{Utama:2015hva}, we assume all connection weights to be
independent and adopt a Gaussian prior centered around zero and with
four \emph{hyperparameters} controlling the width of each set of the
four connection weights $\omega\!=\!\{a,b,c,d\}$.  As in
Ref.\,\cite{Neal1996}, we assume a ``gamma'' probability distribution
for each of the associated hyperparameters, properly adjusted so that
it penalizes a large spread in the weights. For further details on
methods for determining hyperparameters, see
Refs.\,\cite{Mackay:1995,Mackay:1999}.


\section{Results}
\label{Results}

Having set up the formalism that will be implemented to describe
and predict nuclear charge radii, we are now in a position to discuss
our results.  Predictions for charge radii will be made using a
variety of macroscopic and microscopic models that will then be
refined through Bayesian training of neural networks. In most of these
cases our predictions will be compared against experimental results
using the compilations by Angeli and
collaborators\,\cite{Angeli:2004,Angeli:2013}. Specifically, we start
with an underlying nuclear model that provides predictions for charge
radii all throughout the nuclear chart. Next, we refine the nuclear
model predictions by focusing our attention on the residuals between
the theoretical predictions and the experimental data. That is, the
Bayesian learning is applied not to the overall predictions of the
nuclear model but to the discrepancy.  Namely, the target
function defining the likelihood function in Eq.\,(\ref{likelihood})
is given by
\begin{equation} 
  t(x) = \delta R_{\rm ch}(x) = 
  R_{\rm ch}^{\rm (exp)}(x)\!-\!R_{\rm ch}^{\rm (th)}(x).
 \label{residuals}
\end{equation}
In essence, our aim in this two-prong approach is to include as much
physics as possible in the underlying nuclear model and then hope that
the Bayesian refinement will adequately capture most of the physics
that is missing from the model\,\cite{Utama:2015hva}.

Once the theoretical predictions from the nuclear model have been
generated, the BNN refinement starts by separating the available
experimental data into two disjoint sets: a learning set and a
validation set. The learning set consists of a subset of nuclei
contained within the experimental database that will be used to train
the network, {\sl i.e.,} to determine the connection weights (or
parameters) defined in Eq.\,(\ref{ANN}). In contrast, the validation
set comprises those other nuclei that, while still in the existent
experimental database, were not used in the training of the
network. If charge radii within the validation set are accurately
reproduced, then one provides predictions for the charge radii of
nuclei that have not yet been measured but are of particular interest
to other applications or whose measurement may be critical in
constraining nuclear models.

As of 2011, the available data set of experimental charge radii
consisted of 820 nuclei beyond ${}^{40}$Ca ({\sl i.e.,} with
$Z\!\ge\!20$ and $A\!\ge\!40$)\,\cite{Angeli:2013}.  We decided to
ignore light nuclei because both the macroscopic and microscopic
nuclear models used in this work are not particularly suitable to
describe such region of the nuclear chart. For charge radii, the
division between the learning set and the validation set follows the
chronological progression of the compilations by Angeli and
collaborators. That is, for the learning set we use the 722 nuclei
(beyond ${}^{40}$Ca) available in the earlier 2004
compilation\,\cite{Angeli:2004}. We then use the additional 98 new
measurements reported in the latest compilation\,\cite{Angeli:2013} as
the validation set. It is important to note that this division between
learning and validation sets differs from the one adopted in our
earlier work on nuclear masses\,\cite{Utama:2015hva}.  There, the
selection of nuclei belonging to each set was done at random; this
procedure will also be adopted later in the paper. Now, however,
we immediately test the suitability of BNN to extrapolate into
unknown regions of the nuclear chart. Other than this, the training of
the network follows closely the procedure adopted previously. That is,
with two input variables ($Z$ and $A$) and $H\!=\!40$ hidden nodes, a
total of $1\!+\!4H\!=\!161$ weights must be calibrated.  To do so, we
use the \emph{Flexible Bayesian Modeling} package by Neal described in
detail in Ref.\,\cite{Neal1996}. After an initial thermalization phase
consisting of 500 sets, we accumulate data for a total of 100
iterations that are then used to determine the statistical properties
of the charge radius of various nuclei, such as their averages and
variances as in Eqs.\,(\ref{Avgfn}) and (\ref{Errorfn}). Finally, to
assess the quality of the BNN refinement, {\sl i.e.,} the resulting
neural network function $f(x,\omega)$, we compute the root-mean-square (rms)
deviation between the theoretical predictions and the experimental
data as
\begin{equation}
 \sigma_{\rm rms}^{2} = \frac{1}{X_{\rm v}}\sum_{x=1}^{X_{\rm v}} 
 \Big[R_{\rm ch}^{\rm (exp)}(x)\!-\!R_{\rm ch}^{\rm (th)}(x)\Big]^{2},
 \label{MSdeviation}
\end{equation}
where in this case $X_{\rm v}\!=\!98$ is the total number of nuclei in 
the validation set. 

We now proceed to tabulate the rms deviation as per
Eq.\,(\ref{MSdeviation}) for a representative set of both mic-mac and
purely microscopic models. Among the former, we include the ELD 
model of Myers\,\cite{Myers:1983} as well as the predictions from 
Zhang and collaborators\,\cite{Zhang:2002} with the more accurate
parametrization by Bayram {\sl et al.}\,\cite{Bayram:2013rad}; see 
Eqs.\,(\ref{ELD}) and (\ref{Zhang4}). As for the microscopic models, 
we rely on three accurately-calibrated relativistic energy density functionals, 
namely, NL3\,\cite{Lalazissis:1996rd}, FSUGold\,\cite{Todd-Rutel:2005fa}, 
and FSUGarnet\,\cite{Chen:2014mza}. The model parameters for these
three relativistic density functionals, as per Eq.\,(\ref{LDensity}), are 
tabulated in Table\,\ref{Table1}. It is important to note that in addition to 
relativistic energy density functionals, \emph{non-relativistic} models based 
on Skyrme interactions have been very successful in the description of 
charge radii; see for example Refs.\,\cite{Richter:2003wi,Kortelainen:2010hv,
Erler:2012qd} and references contained therein.
\begin{widetext}
\begin{center}
\begin{table}[h]
\begin{tabular}{|l||c|c|c|c|c|c|c|c|}
\hline\rule{0pt}{2.5ex}   
\!\!Model   &  $m_{\rm s}$  &  $g_{\rm s}^2$  &  $g_{\rm v}^2$  &  $g_{\rho}^2$  
                  &  $\kappa$       &  $\lambda$    &  $\zeta$       &   $\Lambda_{\rm v}$  \\
\hline
\hline
NL3              & 508.194000 & 104.387100 & 165.585400 &  79.600000 & 3.859900  & $-$0.015905  & 0.000000  & 0.000000  \\
FSUGold     & 491.500000 & 112.199551 & 204.546943 & 138.470113 & 1.420333  & $+$0.023762 & 0.060000  & 0.030000  \\
FSUGarnet  & 496.939473 & 110.349189 & 187.694676 & 192.927428 & 3.260179  & $-$0.003551 & 0.023499  & 0.043377   \\ 
\hline
\end{tabular}
\caption{Model parameters for the accurately-calibrated relativistic
energy density functionals used in this work. The models included are:
NL3\,\cite{Lalazissis:1996rd}, FSUGold\,\cite{Todd-Rutel:2005fa}, and
FSUGarnet\,\cite{Chen:2014mza}. The parameter $\kappa$ 
and all the masses are given in MeV.  In all cases the omega-meson, 
rho-meson, and nucleon masses have been fixed at $m_{\rm v}\!=\!782.5$,
$m_{\rho}\!=\!763$, and $M\!=\!939$, respectively.}
\label{Table1}
\end{table}
\end{center}
\end{widetext}

We display in Table\,\ref{Table2} the rms deviation as predicted by
the above set of models for the charge radius of all nuclei beyond
${}^{40}$Ca ($Z\!\ge\!20$ and $A\!\ge\!40$) that appear in the latest
compilation by Angeli and Marinova\,\cite{Angeli:2013}. The table
displays results for: (a) the \emph{learning set}, consisting of the
charge radii of the 722 nuclei tabulated in an earlier
compilation\,\cite{Angeli:2004}; (b) the \emph{validation set}, that
includes the 98 additional nuclei appearing in the latest compilation;
and (c) the \emph{entire set} of 820 nuclei. We observe that the
``raw" predictions ({\sl i.e.,} before BNN refinement) generate rms
deviations $\sigma_{\rm pre}$ of the order of several one-hundredth
fermis. Yet, as expected from a model that only depends on the mass
number $A$, the worst performance is displayed by the ELD model.
However, once properly trained, the improvement is truly impressive
(see $\sigma_{\rm post}$) as it rivals the predictions of some of the
most sophisticated models available to date. Even though the
improvement is indeed impressive, ELD fails to capture the basic
essence of our philosophy, namely, a raw model that includes as much
physics as possible which is then fine tuned through BNN
refinement.  In this sense, the microscopic models conform better to
this paradigm. Whereas all the models show consistent improvements
after BNN refinement, the microscopic models see the smallest
change. This suggests that the parameters of these models
properly encode essential features of the nuclear dynamics. It is
important to note, however, that in calibrating these density
functionals, only a handful of charge radii were included in the
fit\,\cite{Todd-Rutel:2005fa,Chen:2014sca,Chen:2014mza}.

\begin{center}
\begin{table}[h]

\textbf{Learning Set}
\\
\begin{tabular}{|l||c|c|c|c|c|}
 \hline
 Model & ELD & Zhang & NL3 & FSUGold & FSUGarnet \\
 \hline
 $\sigma_{\rm pre}$\,(fm)   & 0.0628 & 0.0384 & 0.0304 & 0.0336 & 0.0373   \\
$\sigma_{\rm post}$\,(fm)   & 0.0210 & 0.0207 & 0.0223 & 0.0214 & 0.0215  \\
\hline
$\Delta\sigma/\sigma_{\rm pre}$  & 0.67 & 0.46 & 0.27 & 0.36 & 0.42  \\

\hline
\end{tabular}
\vspace{0.5cm}
\\
\textbf{Validation Set}
\\
\begin{tabular}{|l||c|c|c|c|c|}
 \hline
Model & ELD &  Zhang & NL3 & FSUGold & FSUGarnet \\
 \hline
 $\sigma_{\rm pre}$\,(fm)   & 0.0595  & 0.0461 & 0.0412 & 0.0432 &  0.0510   \\
$\sigma_{\rm post}$\,(fm)   & 0.0262 & 0.0280 & 0.0280 & 0.0298 &  0.0327  \\
\hline
$\Delta\sigma/\sigma_{\rm pre}$    & 0.56 & 0.39 & 0.32 & 0.31 & 0.36  \\

\hline
\end{tabular}
\vspace{0.5cm}
\\
\textbf{Entire Set}
\\
\begin{tabular}{|l||c|c|c|c|c|}
 \hline
 Model & ELD & Zhang & NL3 & FSUGold & FSUGarnet \\
 \hline
 $\sigma_{\rm pre}$\,(fm)   & 0.0639  & 0.0394 & 0.0319 & 0.0349 &  0.0392   \\
$\sigma_{\rm post}$\,(fm)   & 0.0217 & 0.0217 & 0.0230 & 0.0225 &  0.0231  \\
\hline
$\Delta\sigma/\sigma_{\rm pre}$    & 0.66 & 0.45 & 0.28 & 0.36 & 0.41  \\

\hline
\end{tabular}

\caption{Root-mean-square deviation as predicted by a representative set of 
models for the charge radii of 722 nuclei (the learning set), 98 nuclei (the 
validation set), and 820 nuclei (the entire set); see text for details.}
\label{Table2}
\end{table}
\end{center}

\begin{figure}[h]
\vspace{-0.05in}
\includegraphics[width=0.6\columnwidth,angle=0]{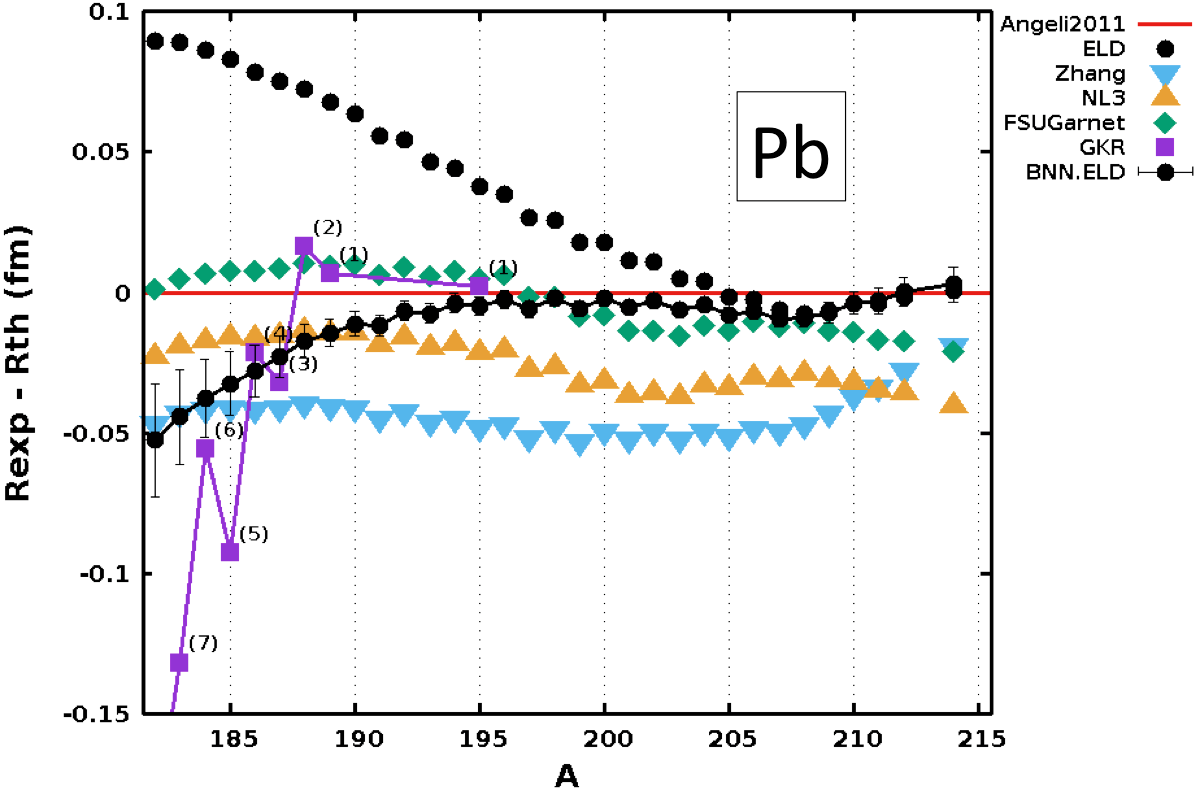}
\caption{Bare predictions for the charge radii of the lead 
isotopes (Z$=$82) relative to experiment\,\cite{Angeli:2013} 
for some of the models considered in the text. Also shown 
are predictions from the Garvey-Kelson relations using an 
iterative method to extrapolate farther into the neutron-deficient
 region. Finally, denoted with error bars are the 
BNN-improved predictions of the extended liquid drop 
model.} 
\label{Fig2}
\end{figure}

Having addressed the impact of the BNN refinement across the full
nuclear chart, we turn our focus to a few of the isotopic chains that
have seen the largest experimental gain in transitioning from the
2004\,\cite{Angeli:2004} to the latest\,\cite{Angeli:2013}
compilation; these are the most abundant isotopes in the validation
set. In particular, the isotopic chains in yttrium ($Z\!=\!39$), lead
($Z\!=\!82$), and bismuth ($Z\!=\!83$) have benefited greatly from the
remarkable experimental progress in laser spectroscopy.  Indeed, our
knowledge of charge radii in these isotopes has expanded from 1 to 16,
from 23 to 32, and from 1 to 12 nuclei, respectively. As an
illustration, we display in Fig.\,\ref{Fig2} the difference between
theory and experiment for the charge radii of lead that, with the
exception of ${}^{213}$Pb, are presently known from ${}^{182}$Pb all
the way to ${}^{214}$Pb\,\cite{Angeli:2013}.  Shown in the plot are
bare predictions from two mic-mac models (ELD and Zhang) and two
relativistic density functionals (NL3 and FSUGarnet). Not
surprisingly, among these the ELD predictions show the largest
deviations relative to experiment; the other three models are, at
worst, within $0.05$\,fm of the experimental data along the whole
isotopic chain. Also shown in Fig.\,\ref{Fig2} are predictions using
the Garvey-Kelson relations\,\cite{Piekarewicz:2009av}; see
Table\,\ref{Table3}. Note that the GKR predictions were made
\emph{before} the publication of the experimental results.  We observe
that when enough information on the nearest neighbors to the nucleus
of interest is available, then the GKR predictions are extremely
accurate. However, the \emph{local} GKRs extrapolate very poorly. This
behavior is clearly indicated in the figure as one aims to predict the
charge radii of progressively more neutron-deficient isotopes. To our
knowledge, the only way that such extrapolations can be implemented is
through an iterative procedure that uses ``first-order" GKR
predictions to extrapolate to ``second order'', and so on. For
example, we found enough neighbors in the 2004 compilation to be able
to predict the charge radius of ${}^{189}$Pb. However, at that time it
was not possible to predict the charge radius of ${}^{188}$Pb. Hence,
in order to make a prediction for ${}^{188}$Pb, we are forced to use
the charge radius of ${}^{189}$Pb---which itself was computed using
the GKRs; this is the meaning of the number enclosed in
parenthesis in Fig.\,\ref{Fig2}. Indeed, in order to provide a GKR
prediction for ${}^{182}$Pb ``eight" iterations were needed.  Not
surprisingly, the local Garvey-Kelson relations are highly inaccurate
far away from their region of applicability. In contrast, the improvement
of the modest ELD model after the BNN refinement is very significant. 
These results indicate that, although extrapolations are always risky,
global methods such as BNN are far superior in extrapolating 
than local methods, such as the GKRs that are inherently prone to 
error propagation.

\begin{widetext}
\begin{center}
\begin{table}[h]
\begin{tabular}{|c|c|c|c|c|c|c|c|}
\hline\rule{0pt}{2.5ex}  
 $\!\!Z$ & $N$ & $A$ & Zhang & NL3 & FSUGarnet & GKR & Experiment \\
 \hline
 \hline
  Y & 48 & 87 & 4.2811 & 4.2546 & 4.2305 & 4.2508 & 4.2498(22) \\ 
     & 49 & 88 & 4.2885 & 4.2565 & 4.2325 & 4.2450 & 4.2441(21) \\ 
     & 51 & 90 & 4.3034 & 4.2659 & 4.2442 & 4.2611 & 4.2573(26) \\
 \hline
  Pb & 107 & 189 & 5.4587 & 5.4322 & 5.4080 & 5.4122 & 5.4177(24) \\ 
       & 113  & 195 & 5.4871 & 5.4602 & 5.4338 & 5.4367 & 5.4389(45) \\   
 \hline
  Bi & 124 & 207 & 5.5540 & 5.5377 & 5.5176 & 5.5098 & 5.5103(32) \\ 
      & 125 & 208 & 5.5587 & 5.5424 & 5.5232 & 5.5144 & 5.5147(28) \\  
\hline
\end{tabular}
\caption{Comparison between the bare predictions of a few models
and the corresponding estimates using the Garvey-Kelson relations
for some nuclei that were recently measured\,\cite{Angeli:2013}. The
GKR predictions compare extremely well against experiment but
extrapolate very poorly; see Fig.\,\ref{Fig2}.}
\label{Table3}
\end{table}
\end{center}
\end{widetext}

Having illustrated some of the central features of the BNN approach
using Angeli's 2004 compilation as the learning set and the nearly
hundred new measurements as the validation set, we now turn to the
slightly different strategy adopted in our earlier work that captures the 
robustness of the method. Using the entire data set from the latest 
compilation of 820 nuclei above $^{40}$Ca\,\cite{Angeli:2013}, we 
randomly select 80$\%$ of the nuclei as the learning set and then 
validate the resulting neural network model against the remaining 
20\%.  Unlike the previous results that highlight the 
\emph{extrapolation} properties of the BNN refinement, 
Table \,\ref{Table4} illustrates how well it \emph{interpolates}. 
Qualitatively, no major trend differences are observed between the 
results shown in Tables\,\ref{Table2} and\,\ref{Table4}. However, 
the quantitative improvement after the refinement is better in the 
present case, as extrapolating is always riskier than interpolating.

\begin{center}
\begin{table}[h]

\textbf{Learning Set}
\\
\begin{tabular}{|l||c|c|c|c|c|}
 \hline
Model & ELD &  Zhang & NL3 & FSUGold & FSUGarnet \\
 \hline
 $\sigma_{\rm pre}$\,(fm)   & 0.0645 &  0.0396 & 0.0319 & 0.0352 &  0.0393   \\
$\sigma_{\rm post}$\,(fm)   & 0.0180 &  0.0171 & 0.0187 & 0.0179 &  0.0190  \\
\hline
$\Delta\sigma/\sigma_{\rm pre}$    & 0.72 & 0.57 & 0.41 & 0.49 & 0.52  \\

\hline
\end{tabular}
\vspace{0.5cm}
\\
\textbf{Validation Set}
\\
\begin{tabular}{|l||c|c|c|c|c|}
 \hline
Model & ELD &  Zhang & NL3 & FSUGold & FSUGarnet \\
 \hline
 $\sigma_{\rm pre}$\,(fm)   & 0.0618 &  0.0385 & 0.0318 & 0.0339 &  0.0388   \\
$\sigma_{\rm post}$\,(fm)   & 0.0173 &  0.0163 & 0.0184 & 0.0179 &  0.0173  \\
\hline
$\Delta\sigma/\sigma_{\rm pre}$    & 0.72  & 0.58 & 0.42 & 0.47 & 0.55  \\

\hline
\end{tabular}
\vspace{0.5cm}
\\
\textbf{Entire Set}
\\
\begin{tabular}{|l||c|c|c|c|c|}
 \hline
Model & ELD &  Zhang & NL3 & FSUGold & FSUGarnet \\
 \hline
$\sigma_{\rm pre}$\,(fm)    & 0.0639 &  0.0394 & 0.0319 & 0.0349 &  0.0392   \\
$\sigma_{\rm post}$\,(fm)   & 0.0179 &  0.0169 & 0.0186 & 0.0179 &  0.0187  \\
\hline
$\Delta\sigma/\sigma_{\rm pre}$    & 0.72 &  0.57 & 0.42 & 0.49 & 0.52  \\

\hline
\end{tabular}

\caption{Root-mean-square deviation as predicted by a representative 
set of models for the charge radii of 656 nuclei (the learning set), 164 
nuclei (the validation set), and 820 nuclei (the entire set); see text for 
details.}
\label{Table4}
\end{table}
\end{center}

\begin{figure}[h]
\vspace{-0.05in}
\includegraphics[width=0.7\columnwidth]{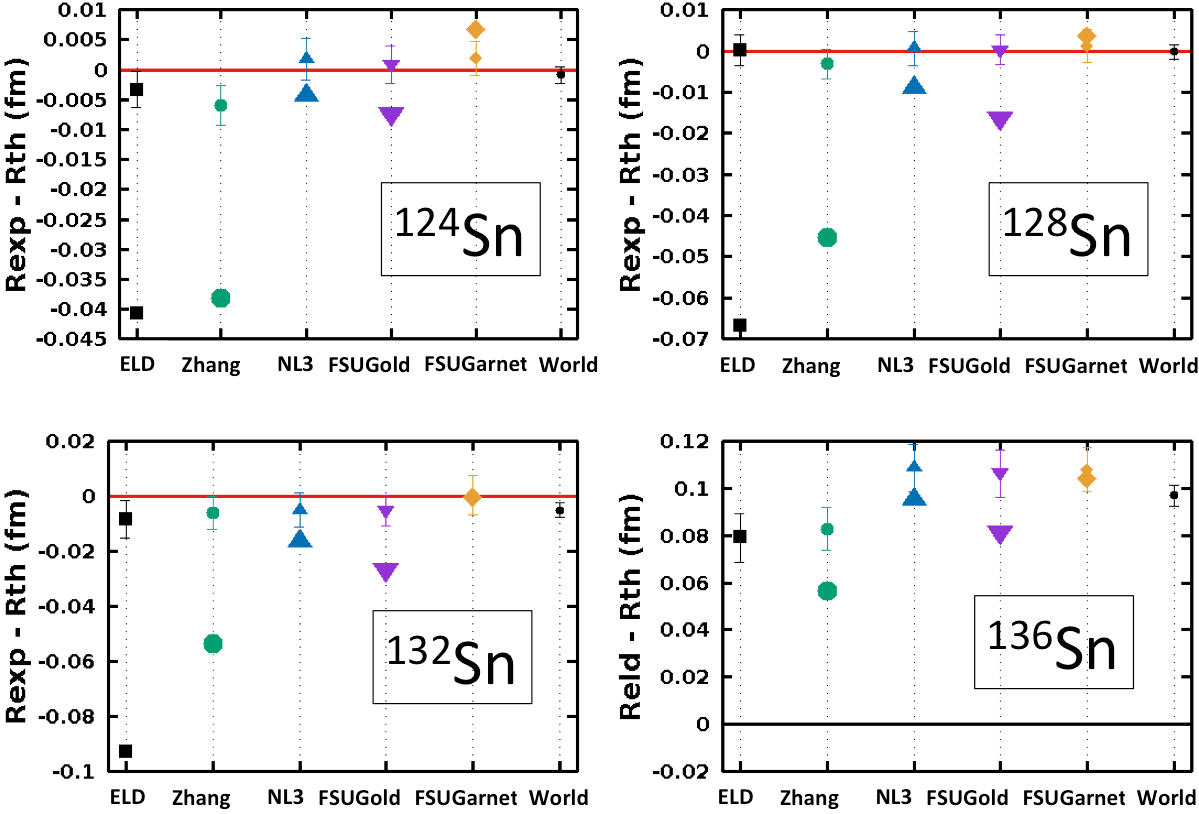}
\caption{Predictions for the charge radii of a few neutron-rich tin 
isotopes from the bare models (without errors bars) and 
after BNN refinement (with error bars). The ``World" predictions
represent a simple way of combining the results of all models
as per Eq.\,(\ref{WorldAvg}). The experimental data are 
from\,\cite{Angeli:2013}, except for ${}^{136}$Sn where the 
bare prediction from the ELD model is adopted as reference.}
\label{Fig3}
\end{figure}

Ultimately, our main goal is to build a model that can \emph{predict}
charge radii which have never been measured. To do so, we use as the
learning set the entire experimental data set containing 820 charge
radii above $^{40}$Ca\,\cite{Angeli:2013}.  All the results presented
hereafter will use a neural network function constructed from all 820
nuclei. We have found that the results obtained in this manner are
both consistent and indeed very similar to those reported in
Table\,\ref{Table4}.  In particular, we conclude that the excellent
agreement with experiment (within $0.02$\,fm) rivals---and
likely exceeds---the best models currently available in the
literature. Finally, by treating the various predictions as statistically 
independent, we can properly combine them into a ``world average''. 
That is,
\begin{equation}
  R_{\rm world}=\sum_{n} \omega_{n}\,R_{n}\,, 
   \hspace{8pt}
  V_{\rm world}=\sum_{n} \omega_{n}^{2}\,V_{n}\,, 
   \hspace{4pt} {\rm and} \hspace{6pt}  
  \omega_{n} = \frac{V_{n}^{-1}}{\sum_{k}V_{k}^{-1}}\,,
 \label{WorldAvg} 
\end{equation}
where the sum runs over all five models and $V_{n}$ represents the
variance of each model. For example, world-average predictions for a
few neutron-rich tin isotopes are displayed in Fig.\,\ref{Fig3},
alongside the corresponding predictions from each individual model.
With the exception of ${}^{136}$Sn, all models are compared against
existing experimental data. In the particular case of ${}^{136}$Sn
where data is not yet available, we use the bare ELD prediction of
$R_{\rm ch}^{\rm {ELD}}\!=\!4.846\,{\rm fm}$ as a baseline. For
reference, FSUGarnet predicts after BNN refinement a charge radius of
about 0.1\,fm smaller than the ELD reference, or 
$R_{\rm ch}\!=\!(4.737\pm0.009)\,{\rm fm}$. We observe that in all 
three cases where experimental measurements are available, the 
BNN refinement leads to significantly reduced scattering, an 
improvement in the theoretical predictions, especially in the case 
of the mic-mac models and ultimately, to world-average predictions 
that are both accurate and precise. This lends credence to our 
approach and validates our prediction of the charge radius of 
${}^{136}$Sn.

\begin{figure}[h]
\vspace{-0.05in}
\includegraphics[width=0.45\columnwidth]{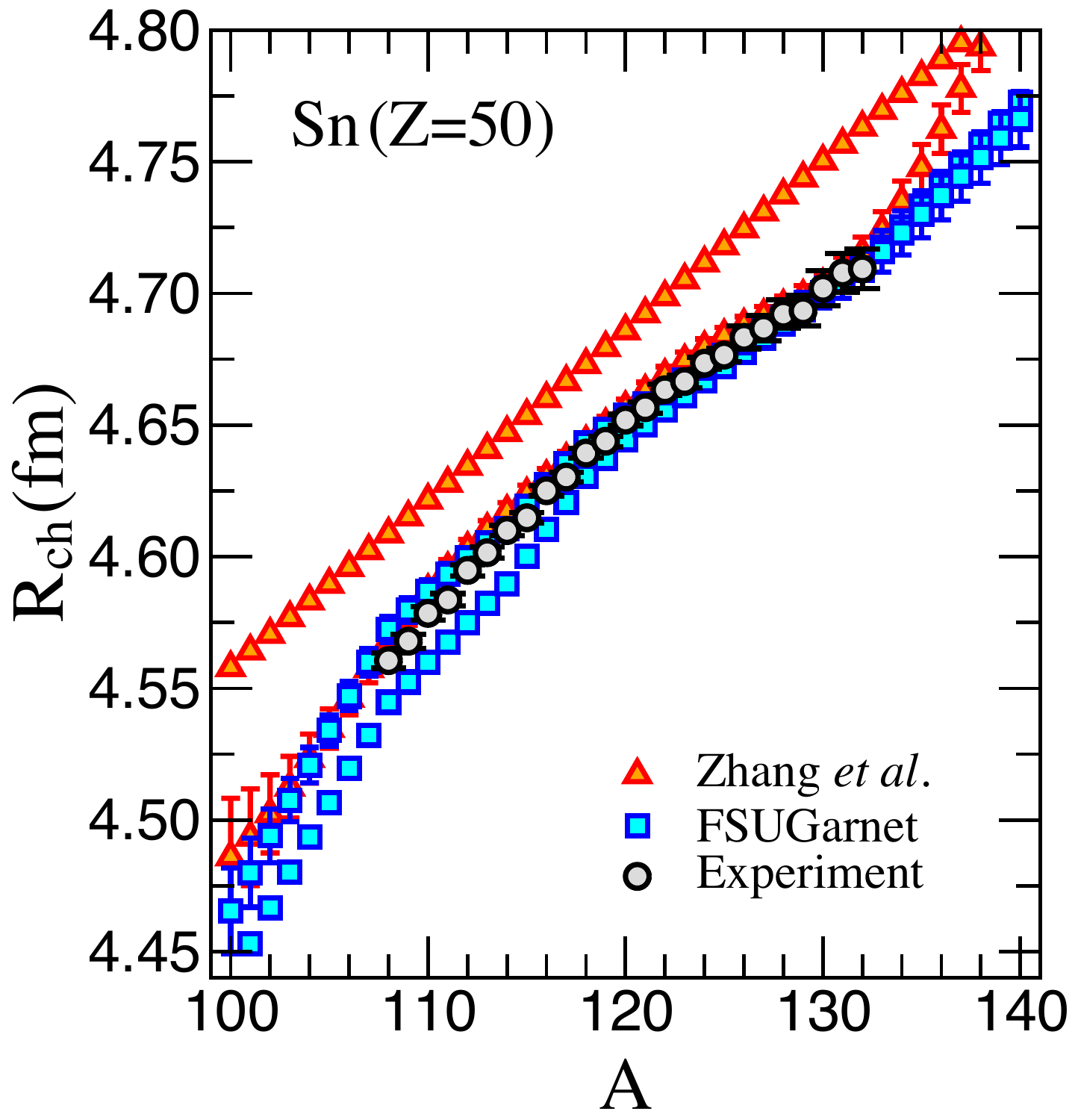}
\caption{Predictions for the charge radii of the tin isotopes using the 
mic-mac model of Zhang {\sl et al.}\,\cite{Zhang:2002,Bayram:2013rad} 
and FSUGarnet\,\cite{Chen:2014mza}. Predictions are displayed without 
errors bars for the bare models and with error bars after the BNN refinement. 
Experimental data---available from ${}^{108}$Sn all the way to 
${}^{132}$Sn---are from Ref.\,\cite{Angeli:2013}.}
\label{Fig4}
\end{figure}

\begin{figure}[h]
\vspace{-0.05in}
\includegraphics[width=0.475\columnwidth]{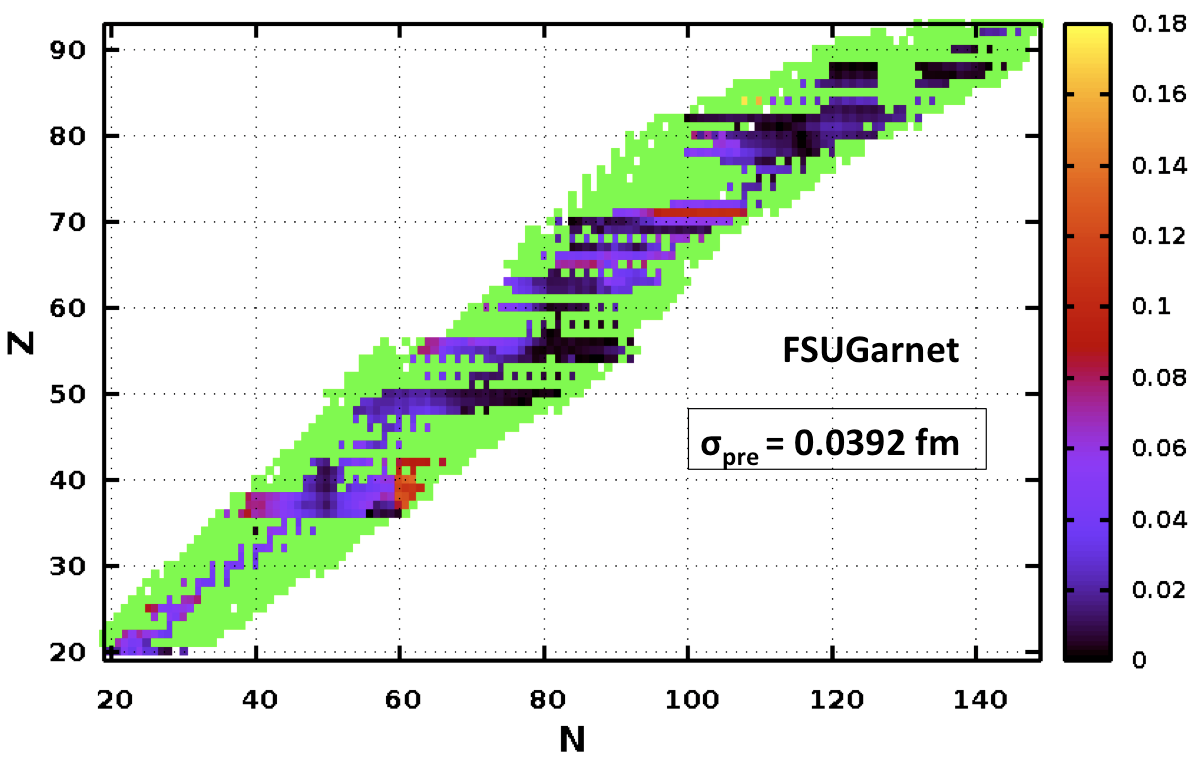}
\hspace{10pt}
\includegraphics[width=0.475\columnwidth]{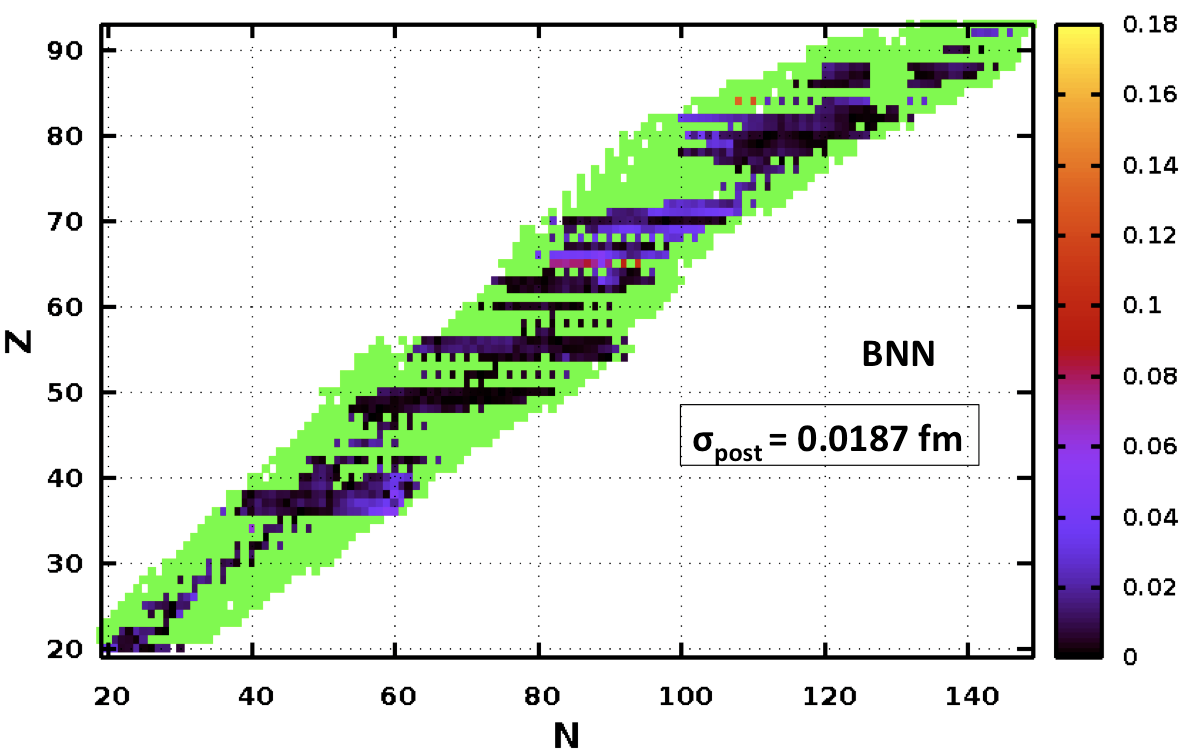}
\caption{FSUGarnet predictions, before and after BNN refinement, 
for the charge radii of all nuclei whose mass is experimentally 
known\,\cite{AME:2012}. Color coded in green are predictions for
those nuclei whose charge radius is presently unknown. Tabulated
values for all these nuclei appear in the supplemental material to
this paper.}
\label{Fig5}
\end{figure}

Having illustrated the systematic improvement in our
prediction for a few isotopes, we now display in Fig.\,\ref{Fig4}
predictions along the whole isotopic chain in tin using the mic-mac
model of Zhang {\sl et al.}\,\cite{Zhang:2002} (with the parameters
from \,\cite{Bayram:2013rad}) and FSUGarnet\,\cite{Chen:2014mza} 
as a representative set of our predictions.  It is worth highlighting the
remarkable experimental progress that has been made in the last few
years in extending the measurement of charge radii far away  from
stability. In the particular case of tin, high precision data is now
available from ${}^{108}$Sn all the way to ${}^{132}$Sn. In many 
ways, Fig.\,\ref{Fig4} captures the true essence of our theoretical
approach: an accurately-calibrated density functional which by
incorporating as much physics as possible provides an excellent
description of the experimental data that is then improved even
further after the BNN refinement. And whereas our aim is for BNN 
to provide a fine tuning, the figure illustrates that even when there 
is significant disagreement between the theoretical predictions and 
experiment, as in the case of the mic-mac model, the approach is 
powerful enough to overcome most of these deficiencies. Utilizing 
this combined DFT+BNN approach (labeled ``Entire Set" in 
Table\,\ref{Table4}) we provide as a supplement to this paper 
predictions for the charge radii of all nuclei beyond ${}^{40}$Ca 
whose mass is experimentally known\,\cite{AME:2012}. This large
collection of nuclei is also displayed as a heat map in Fig.\,\ref{Fig5}
using both ``pre'' and ``post'' FSUGarnet predictions. 
Note that color coded in green are predictions for those nuclei 
whose charge radius is presently unknown. Yet, given the 
remarkable experimental progress in the determination of 
charge radii of nuclei far away from stability, we trust that 
these predictions will be tested against experiment in the 
coming years.

\begin{figure}[h]
\vspace{-0.05in}
\includegraphics[width=0.5\columnwidth]{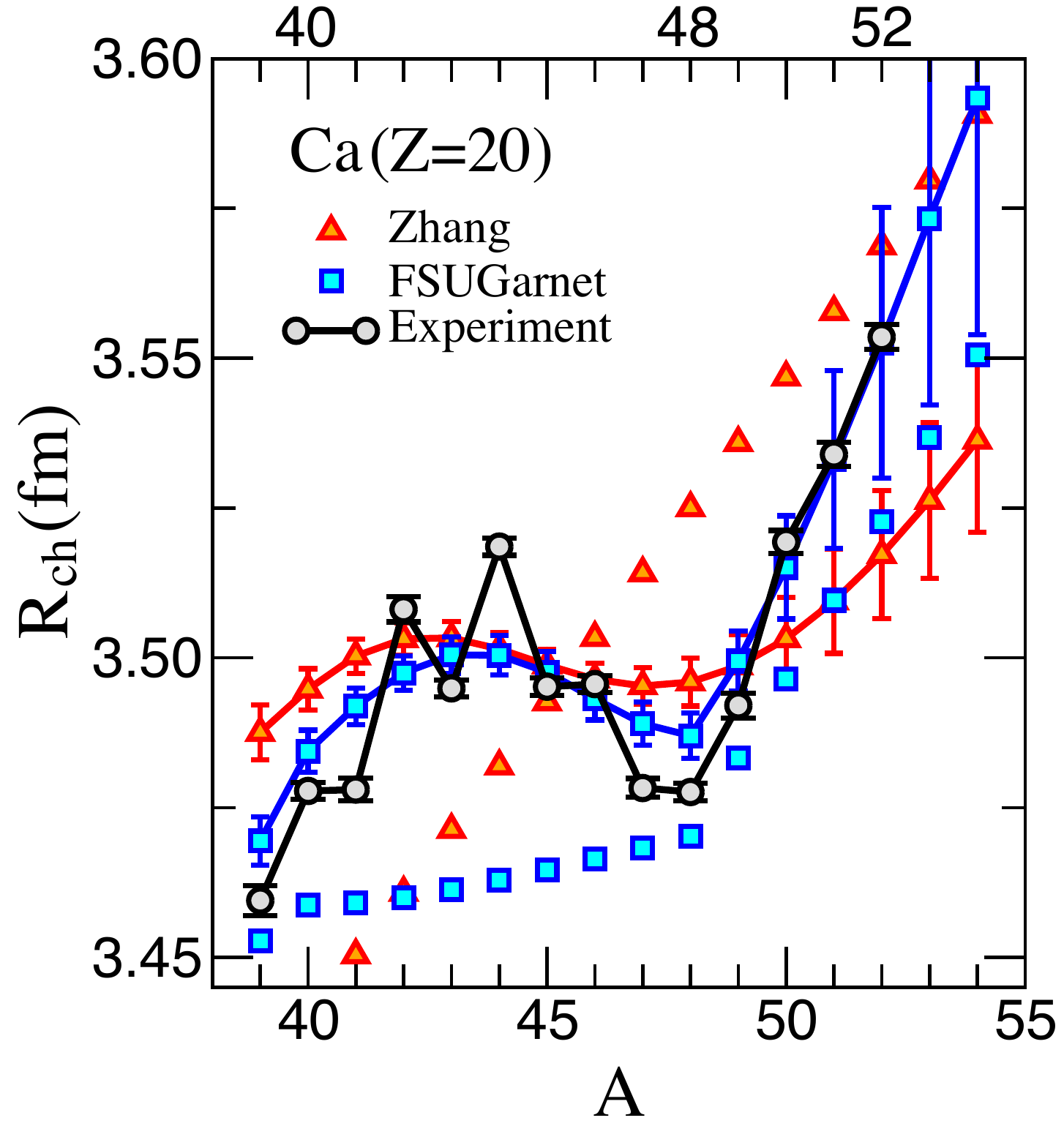}
\caption{Predictions for the charge radii of the calcium isotopes using the 
mic-mac model of Zhang {\sl et al.}\,\cite{Zhang:2002,Bayram:2013rad} 
and FSUGarnet\,\cite{Chen:2014mza}. Predictions are displayed without 
errors bars for the bare models and with error bars after the BNN refinement. 
Experimental data are from Refs.\,\cite{Angeli:2013,Ruiz:2016gne}.}
\label{Fig6}
\end{figure}

Finally, we now return to one of the main motivations behind this
work: understanding the underlying physics behind the intriguing 
trend displayed by the charge radii of the calcium
isotopes\,\cite{Ruiz:2016gne}; see Fig.\,\ref{Fig6}. The evolution 
of the charge radii in the calcium isotopes is puzzling for several
reasons. First, within the region of the stable isotopes, a
long-standing question remains unanswered: why are the charge 
radii of ${}^{40}$Ca and ${}^{48}$Ca practically identical; {\sl i.e.,}
$R_{\rm ch}^{40}\!=\!3.4776(19)\,{\rm fm}$ and 
$R_{\rm ch}^{48}\!=\!3.4771(20)\,{\rm fm}$. Second, although it 
is well known that pairing correlations are behind the pervasive 
odd-even effects observed in a variety of observables throughout 
the nuclear chart, why are these effects so pronounced as the 
system evolves from $N\!=\!20$ to $N\!=\!28$. Finally, as reported 
in Ref.\,\cite{Ruiz:2016gne}, there is a large and unexpected increase 
in the size of the neutron-rich calcium isotopes beyond $N\!=\!28$. 
Unexpected in the sense that theoretical results obtained with either 
ab initio approaches, configuration interaction, or density functional 
theory are unable to reproduce the complex experimental trend. We 
note that some studies using energy density functionals that include 
a sophisticated treatment of pairing
correlations\,\cite{FAYANS:2000,Saperstein:2011} or shell model
calculations using a relatively large model
space\,\cite{Caurier:2001np} appear to successfully reproduce the
charge radii along the isotopic chain in calcium.  Yet, by the
authors' own admission\,\cite{FAYANS:2000}, the \emph{anomalous
behavior of charge radii in {\rm Ca} isotopes was reproduced in
excellent agreement with experiment ... by scaling the pairing
effective interaction which was extracted from the lead chain by a
factor of 1.35}, thereby concluding that \emph{a universal
parametrization of the pairing force is still
lacking}\,\cite{FAYANS:2000}. Similarly, it is concluded in
Ref.\,\cite{Caurier:2001np} that whereas the experimental trends are
well reproduced by shell model calculations, \emph{the magnitude of
the calculated shifts is smaller than the experiment suggests.}

Unfortunately, our own DFT+BNN approach does not fare any better.  In
an effort to eliminate, or at least mitigate, the discrepancy with
experiment, we have modified our fitting strategy in the following
manner.  First, we have limited the training set to the region between
oxygen ($Z\!=\!8$) and neodymium ($Z\!=\!60$). Second, we have biased
our results in favor of the calcium isotopes by readjusting the
weights in the likelihood function; all the weights are enhanced by a
factor of two, except in the case of the two doubly magic nuclei
${}^{40}$Ca and ${}^{48}$Ca where they are enhanced by a factor of
three. Thus, the training of the neural network involves incorporating the
entirety of the 472 nuclei in this
region\,\cite{Angeli:2013}. Finally, keeping all other inputs
unchanged, we re-train the neural network function using only the
models of Zhang and FSUGarnet.

In Fig.\,\ref{Fig6} we show the bare results of the Zhang {\sl et al.}
model (red triangles with no error bars) that predict a nearly linear
dependence with $A$ (or $N$) which in no way resembles the trend
displayed by the data.  In turn, predictions with FSUGarnet (blue
squares with no error bars) show a behavior that is characteristic of
most energy density functionals: a relatively slow increase in the
$A\!=\!40$-48 region preceding a sharp rise beyond ${}^{48}$Ca,
although not as sharp as demanded by the data.  BNN results for both
models, now shown with uncertainty bars, improve significantly after
the refinement, particularly in the case of FSUGarnet that now
accounts, albeit with large error bars, for the steep rise in the data
beyond ${}^{48}$Ca. However, although both models develop some
structure in the $A\!=\!40$-48 region after BNN refinement, neither of
them can reproduce the dramatic odd-even effects displayed by the
data. We must conclude then that at present, the intriguing trend
displayed by the charge radii of the calcium isotopes remains ``a
riddle, wrapped in a mystery, inside an enigma''.


\section{Conclusions}
\label{Conclusions}

The mass and size of atomic nuclei feature among their most
distinctive and fundamental characteristics. Within a few years after
the discovery of the neutron by Chadwick, Bethe and Weizs\"acker
conceived the liquid drop model, which up to this day continues to
provide a qualitative description of both the mass and radius of
nuclei. During the past eight decades, sophisticated
theoretical approaches have been developed to improve this very early
description. Ultimately, the goal of nuclear theory is to develop a
predictive understanding of the structure and dynamics of nuclei based
on QCD, the fundamental theory of the strong interaction. While
significant progress has been made in understanding QCD in the
low-energy regime, at present most of our understanding of nuclear
phenomena derives from effective field theories of nucleons/mesons
that incorporate the underlying symmetries of QCD. Thus, whereas the
predictive capability of nuclear theory has increased dramatically, it
has not yet reach the level of precision that is required to guide
applications into other areas, primarily fundamental symmetries and
astrophysics.  In response to this situation, we have recently
proposed a hybrid approach that combines sophisticated theoretical
models with Bayesian neural networks. The basic tenet of this approach
is to start with a model that incorporates as much physics as possible
which then gets refined by building a Bayesian neural network
model. This approach was recently implemented with considerable
success in the case of nuclear masses of relevance to the crustal
composition of neutron stars\,\cite{Utama:2015hva}.  Motivated by some
recent results along the isotopic chain in
calcium\,\cite{Ruiz:2016gne}, we have extended the approach to
the description of nuclear charge radii.

As illustrated previously for nuclear masses\,\cite{Utama:2015hva},
the hybrid approach starts by invoking both mic-mac and microscopic
models. Among the mic-mac models we have used two simple extensions of
the liquid-drop model, and in the case of the microscopic models we
relied on accurately-calibrated relativistic parametrizations. In all
cases, the theoretical models provide baseline estimates that often
compare favorably against the extensive database of experimental
charge radii\,\cite{Angeli:2013}. However, those models, although
accurate, can all benefit from further refinement.  This additional
refinement is implemented through training a Bayesian neural network
on the \emph{residuals} between the theoretical predictions and
experiment. The BNN formalism is an approximation scheme that relies
on the application of Bayes' theorem and a highly nonlinear neural
network function aiming at refining our predictions. Besides improved
predictions, BNN has the enormous advantage of providing an estimate
of the theoretical uncertainties. Given that neural network learning
involves information on charge radii all throughout the nuclear chart,
we found it pertinent to compare this \emph{global} approach against
predictions from the Garvey-Kelson relations, a highly successful
\emph{local} method.

Several optimization protocols were adopted to test the reliability
of our proposed approach, from one that uses the 2004 compilation
by Angeli\,\cite{Angeli:2004} as the learning set and the new nuclei
appearing in the latest compilation as the validation 
set\,\cite{Angeli:2013}, to one in which the full 2013 dataset is
used to calibrate the neural network function. In all cases we 
saw considerable improvement after the BNN refinement, 
ultimately achieving rms deviations below 0.02\,fm for a collection
of 820 nuclei. Although it is impractical to illustrate the performance
of our approach for many isotopic chains, several important findings
were obtained. First, we showed that whereas local methods such as
 the Garvey-Kelson relations work extremely well when enough
neighbors are available, they extrapolate---in contrast to the global
BNN approach---extremely poorly. Second, we regarded our results 
for the tin isotopes as a ``textbook example" of the combined DFT+BNN
method. Indeed, although we found very good initial agreement using 
the bare predictions of FSUGarnet, these predictions became even better after the BNN 
refinement and were properly 
supplemented with error bars. Finally, like many before us, we failed to reproduce 
in detail the complex behavior of the calcium isotopes. While a 
convincing explanation for the pronounced odd-even effects in the 
$A\!=\!40$-$48$ region has been proven elusive, recent experiments at 
ISOLDE have complicated matters even further 
by reporting a very sharp increase in the charge radii of the neutron-rich
calcium isotopes that is inconsistent with theoretical 
predictions\,\cite{Ruiz:2016gne}. Yet, by a ``proper'' adjustment of 
the neural network function, we were able to eliminate some of 
these discrepancies. Indeed, the steep rise can be accounted 
for within the large theoretical uncertainties. However, the physics
 behind the odd-even staggering and the very 
steep rise remains a mystery.

In summary, BNN training has been shown to be a promising new tool to
refine the predictions of existing nuclear models. In this paper we
have extended our earlier work on nuclear masses to nuclear charge
radii. In cases in which the underlying physics model is sound, then
BNN refinement works extremely well. In contrast, if there is
important physics missing from the underlying model, it is unlikely
that BNN by itself can fix such deficiencies.  That is, although the
BNN framework is very robust for model refinement, an essential
requirement for its success is that the quantity to be fitted, such as
model residuals, be a relatively \emph{smooth} function of the input
parameters. In other words, we are confident that if the physics model is able
to reproduce the main trends in the data, then BNN can provide the
fine-tuning necessary to reach the precision demanded by astrophysics
(and other) applications.


\begin{acknowledgments}
 We are very grateful to Dr. Harrison Prosper and Dr. Michelle Kuchera 
 for many fruitful discussions. This material is based upon work supported 
 by the U.S. Department of Energy Office of Science, Office of Nuclear 
 Physics under Award Number DE-FD05-92ER40750.
\end{acknowledgments}


\bibliography{./BNNonChargeRadii.bbl}

\begin{thebibliography}{66}%
\makeatletter
\providecommand \@ifxundefined [1]{%
 \@ifx{#1\undefined}
}%
\providecommand \@ifnum [1]{%
 \ifnum #1\expandafter \@firstoftwo
 \else \expandafter \@secondoftwo
 \fi
}%
\providecommand \@ifx [1]{%
 \ifx #1\expandafter \@firstoftwo
 \else \expandafter \@secondoftwo
 \fi
}%
\providecommand \natexlab [1]{#1}%
\providecommand \enquote  [1]{``#1''}%
\providecommand \bibnamefont  [1]{#1}%
\providecommand \bibfnamefont [1]{#1}%
\providecommand \citenamefont [1]{#1}%
\providecommand \href@noop [0]{\@secondoftwo}%
\providecommand \href [0]{\begingroup \@sanitize@url \@href}%
\providecommand \@href[1]{\@@startlink{#1}\@@href}%
\providecommand \@@href[1]{\endgroup#1\@@endlink}%
\providecommand \@sanitize@url [0]{\catcode `\\12\catcode `\$12\catcode
  `\&12\catcode `\#12\catcode `\^12\catcode `\_12\catcode `\%12\relax}%
\providecommand \@@startlink[1]{}%
\providecommand \@@endlink[0]{}%
\providecommand \url  [0]{\begingroup\@sanitize@url \@url }%
\providecommand \@url [1]{\endgroup\@href {#1}{\urlprefix }}%
\providecommand \urlprefix  [0]{URL }%
\providecommand \Eprint [0]{\href }%
\providecommand \doibase [0]{http://dx.doi.org/}%
\providecommand \selectlanguage [0]{\@gobble}%
\providecommand \bibinfo  [0]{\@secondoftwo}%
\providecommand \bibfield  [0]{\@secondoftwo}%
\providecommand \translation [1]{[#1]}%
\providecommand \BibitemOpen [0]{}%
\providecommand \bibitemStop [0]{}%
\providecommand \bibitemNoStop [0]{.\EOS\space}%
\providecommand \EOS [0]{\spacefactor3000\relax}%
\providecommand \BibitemShut  [1]{\csname bibitem#1\endcsname}%
\let\auto@bib@innerbib\@empty
\bibitem [{\citenamefont {von Weizs{\"a}cker}(1935)}]{Weizsacker:1935}%
  \BibitemOpen
  \bibfield  {author} {\bibinfo {author} {\bibfnamefont {C.~F.}\ \bibnamefont
  {von Weizs{\"a}cker}},\ }\href@noop {} {\bibfield  {journal} {\bibinfo
  {journal} {Z. Physik}\ }\textbf {\bibinfo {volume} {96}},\ \bibinfo {pages}
  {431} (\bibinfo {year} {1935})}\BibitemShut {NoStop}%
\bibitem [{\citenamefont {Bethe}\ and\ \citenamefont
  {Bacher}(1936)}]{Bethe:1936}%
  \BibitemOpen
  \bibfield  {author} {\bibinfo {author} {\bibfnamefont {H.~A.}\ \bibnamefont
  {Bethe}}\ and\ \bibinfo {author} {\bibfnamefont {R.~F.}\ \bibnamefont
  {Bacher}},\ }\href {\doibase 10.1103/RevModPhys.8.82} {\bibfield  {journal}
  {\bibinfo  {journal} {Rev. Mod. Phys.}\ }\textbf {\bibinfo {volume} {8}},\
  \bibinfo {pages} {82} (\bibinfo {year} {1936})}\BibitemShut {NoStop}%
\bibitem [{\citenamefont {M\"oller}\ and\ \citenamefont
  {Nix}(1981)}]{Moller:1981zz}%
  \BibitemOpen
  \bibfield  {author} {\bibinfo {author} {\bibfnamefont {P.}~\bibnamefont
  {M\"oller}}\ and\ \bibinfo {author} {\bibfnamefont {J.~R.}\ \bibnamefont
  {Nix}},\ }\href@noop {} {\bibfield  {journal} {\bibinfo  {journal} {Atom.
  Data Nucl. Data Tabl.}\ }\textbf {\bibinfo {volume} {26}},\ \bibinfo {pages}
  {165} (\bibinfo {year} {1981})}\BibitemShut {NoStop}%
\bibitem [{\citenamefont {M\"oller}\ and\ \citenamefont
  {Nix}(1988)}]{Moller:1988}%
  \BibitemOpen
  \bibfield  {author} {\bibinfo {author} {\bibfnamefont {P.}~\bibnamefont
  {M\"oller}}\ and\ \bibinfo {author} {\bibfnamefont {J.~R.}\ \bibnamefont
  {Nix}},\ }\href@noop {} {\bibfield  {journal} {\bibinfo  {journal} {Atom.
  Data Nucl. Data Tabl.}\ }\textbf {\bibinfo {volume} {39}},\ \bibinfo {pages}
  {213} (\bibinfo {year} {1988})}\BibitemShut {NoStop}%
\bibitem [{\citenamefont {M\"oller}\ \emph {et~al.}(1995)\citenamefont
  {M\"oller}, \citenamefont {Nix}, \citenamefont {Myers},\ and\ \citenamefont
  {Swiatecki}}]{Moller:1993ed}%
  \BibitemOpen
  \bibfield  {author} {\bibinfo {author} {\bibfnamefont {P.}~\bibnamefont
  {M\"oller}}, \bibinfo {author} {\bibfnamefont {J.~R.}\ \bibnamefont {Nix}},
  \bibinfo {author} {\bibfnamefont {W.~D.}\ \bibnamefont {Myers}}, \ and\
  \bibinfo {author} {\bibfnamefont {W.~J.}\ \bibnamefont {Swiatecki}},\
  }\href@noop {} {\bibfield  {journal} {\bibinfo  {journal} {Atom. Data Nucl.
  Data Tabl.}\ }\textbf {\bibinfo {volume} {59}},\ \bibinfo {pages} {185}
  (\bibinfo {year} {1995})}\BibitemShut {NoStop}%
\bibitem [{\citenamefont {Duflo}\ and\ \citenamefont
  {Zuker}(1995)}]{Duflo:1995}%
  \BibitemOpen
  \bibfield  {author} {\bibinfo {author} {\bibfnamefont {J.}~\bibnamefont
  {Duflo}}\ and\ \bibinfo {author} {\bibfnamefont {A.}~\bibnamefont {Zuker}},\
  }\href {\doibase 10.1103/PhysRevC.52.R23} {\bibfield  {journal} {\bibinfo
  {journal} {Phys. Rev. C}\ }\textbf {\bibinfo {volume} {52}},\ \bibinfo
  {pages} {R23} (\bibinfo {year} {1995})}\BibitemShut {NoStop}%
\bibitem [{\citenamefont {Brown}\ \emph {et~al.}(1984)\citenamefont {Brown},
  \citenamefont {Bronk},\ and\ \citenamefont {Hodgson}}]{Brown:1984}%
  \BibitemOpen
  \bibfield  {author} {\bibinfo {author} {\bibfnamefont {B.~A.}\ \bibnamefont
  {Brown}}, \bibinfo {author} {\bibfnamefont {C.~R.}\ \bibnamefont {Bronk}}, \
  and\ \bibinfo {author} {\bibfnamefont {P.~E.}\ \bibnamefont {Hodgson}},\
  }\href@noop {} {\bibfield  {journal} {\bibinfo  {journal} {J. Phys. G: Nucl.
  Phys.}\ }\textbf {\bibinfo {volume} {10}},\ \bibinfo {pages} {1683} (\bibinfo
  {year} {1984})}\BibitemShut {NoStop}%
\bibitem [{\citenamefont {Myers}\ and\ \citenamefont
  {Schmidt}(1983)}]{Myers:1983}%
  \BibitemOpen
  \bibfield  {author} {\bibinfo {author} {\bibfnamefont {W.}~\bibnamefont
  {Myers}}\ and\ \bibinfo {author} {\bibfnamefont {K.-H.}\ \bibnamefont
  {Schmidt}},\ }\href@noop {} {\bibfield  {journal} {\bibinfo  {journal} {Nucl.
  Phys.}\ }\textbf {\bibinfo {volume} {A410}},\ \bibinfo {pages} {61} (\bibinfo
  {year} {1983})}\BibitemShut {NoStop}%
\bibitem [{\citenamefont {Helm}(1956)}]{Helm:1956zz}%
  \BibitemOpen
  \bibfield  {author} {\bibinfo {author} {\bibfnamefont {R.~H.}\ \bibnamefont
  {Helm}},\ }\href {\doibase 10.1103/PhysRev.104.1466} {\bibfield  {journal}
  {\bibinfo  {journal} {Phys. Rev.}\ }\textbf {\bibinfo {volume} {104}},\
  \bibinfo {pages} {1466} (\bibinfo {year} {1956})}\BibitemShut {NoStop}%
\bibitem [{\citenamefont {Zhang}\ \emph {et~al.}(2002)\citenamefont {Zhang},
  \citenamefont {Meng}, \citenamefont {Zhou},\ and\ \citenamefont
  {Zheng}}]{Zhang:2002}%
  \BibitemOpen
  \bibfield  {author} {\bibinfo {author} {\bibfnamefont {S.~Q.}\ \bibnamefont
  {Zhang}}, \bibinfo {author} {\bibfnamefont {J.}~\bibnamefont {Meng}},
  \bibinfo {author} {\bibfnamefont {S.~G.}\ \bibnamefont {Zhou}}, \ and\
  \bibinfo {author} {\bibfnamefont {J.~Y.}\ \bibnamefont {Zheng}},\ }\href
  {\doibase 10.1016/j.physletb.2013.08.002} {\bibfield  {journal} {\bibinfo
  {journal} {Eur. Phys. J.}\ }\textbf {\bibinfo {volume} {A13}},\ \bibinfo
  {pages} {285} (\bibinfo {year} {2002})}\BibitemShut {NoStop}%
\bibitem [{\citenamefont {Bayram}\ \emph {et~al.}(2013)\citenamefont {Bayram},
  \citenamefont {Akkoyun},\ and\ \citenamefont {Kara}}]{Bayram:2013rad}%
  \BibitemOpen
  \bibfield  {author} {\bibinfo {author} {\bibfnamefont {T.}~\bibnamefont
  {Bayram}}, \bibinfo {author} {\bibfnamefont {S.}~\bibnamefont {Akkoyun}}, \
  and\ \bibinfo {author} {\bibfnamefont {S.~O.}\ \bibnamefont {Kara}},\ }\href
  {\doibase 10.5506/APhysPolB.44.1791} {\bibfield  {journal} {\bibinfo
  {journal} {Acta Physica Polonica B}\ }\textbf {\bibinfo {volume} {44}},\
  \bibinfo {pages} {1791} (\bibinfo {year} {2013})}\BibitemShut {NoStop}%
\bibitem [{\citenamefont {Fayans}\ \emph {et~al.}(2000)\citenamefont {Fayans},
  \citenamefont {Tolokonnikov}, \citenamefont {Trykov},\ and\ \citenamefont
  {Zawischa}}]{FAYANS:2000}%
  \BibitemOpen
  \bibfield  {author} {\bibinfo {author} {\bibfnamefont {S.}~\bibnamefont
  {Fayans}}, \bibinfo {author} {\bibfnamefont {S.}~\bibnamefont
  {Tolokonnikov}}, \bibinfo {author} {\bibfnamefont {E.}~\bibnamefont
  {Trykov}}, \ and\ \bibinfo {author} {\bibfnamefont {D.}~\bibnamefont
  {Zawischa}},\ }\href@noop {} {\bibfield  {journal} {\bibinfo  {journal}
  {Nuclear Physics}\ }\textbf {\bibinfo {volume} {A676}},\ \bibinfo {pages} {49
  } (\bibinfo {year} {2000})}\BibitemShut {NoStop}%
\bibitem [{\citenamefont {Goriely}\ \emph {et~al.}(2010)\citenamefont
  {Goriely}, \citenamefont {Chamel},\ and\ \citenamefont
  {Pearson}}]{Goriely:2010bm}%
  \BibitemOpen
  \bibfield  {author} {\bibinfo {author} {\bibfnamefont {S.}~\bibnamefont
  {Goriely}}, \bibinfo {author} {\bibfnamefont {N.}~\bibnamefont {Chamel}}, \
  and\ \bibinfo {author} {\bibfnamefont {J.}~\bibnamefont {Pearson}},\ }\href
  {\doibase 10.1103/PhysRevC.82.035804} {\bibfield  {journal} {\bibinfo
  {journal} {Phys. Rev.}\ }\textbf {\bibinfo {volume} {C82}},\ \bibinfo {pages}
  {035804} (\bibinfo {year} {2010})}\BibitemShut {NoStop}%
\bibitem [{\citenamefont {Kortelainen}\ \emph {et~al.}()\citenamefont
  {Kortelainen}, \citenamefont {Lesinski}, \citenamefont {More}, \citenamefont
  {Nazarewicz}, \citenamefont {Sarich} \emph {et~al.}}]{Kortelainen:2010hv}%
  \BibitemOpen
  \bibfield  {author} {\bibinfo {author} {\bibfnamefont {M.}~\bibnamefont
  {Kortelainen}}, \bibinfo {author} {\bibfnamefont {T.}~\bibnamefont
  {Lesinski}}, \bibinfo {author} {\bibfnamefont {J.}~\bibnamefont {More}},
  \bibinfo {author} {\bibfnamefont {W.}~\bibnamefont {Nazarewicz}}, \bibinfo
  {author} {\bibfnamefont {J.}~\bibnamefont {Sarich}},  \emph {et~al.},\
  }\href@noop {} {\bibfield  {journal} {\bibinfo  {journal} {Phys.Rev.}\
  }\textbf {\bibinfo {volume} {C82}},\ \bibinfo {pages} {024313}}\BibitemShut
  {NoStop}%
\bibitem [{\citenamefont {Erler}\ \emph {et~al.}(2013)\citenamefont {Erler},
  \citenamefont {Horowitz}, \citenamefont {Nazarewicz}, \citenamefont
  {Rafalski},\ and\ \citenamefont {Reinhard}}]{Erler:2012qd}%
  \BibitemOpen
  \bibfield  {author} {\bibinfo {author} {\bibfnamefont {J.}~\bibnamefont
  {Erler}}, \bibinfo {author} {\bibfnamefont {C.~J.}\ \bibnamefont {Horowitz}},
  \bibinfo {author} {\bibfnamefont {W.}~\bibnamefont {Nazarewicz}}, \bibinfo
  {author} {\bibfnamefont {M.}~\bibnamefont {Rafalski}}, \ and\ \bibinfo
  {author} {\bibfnamefont {P.-G.}\ \bibnamefont {Reinhard}},\ }\href {\doibase
  10.1103/PhysRevC.87.044320} {\bibfield  {journal} {\bibinfo  {journal} {Phys.
  Rev.}\ }\textbf {\bibinfo {volume} {C87}},\ \bibinfo {pages} {044320}
  (\bibinfo {year} {2013})}\BibitemShut {NoStop}%
\bibitem [{\citenamefont {Chen}\ and\ \citenamefont
  {Piekarewicz}(2014)}]{Chen:2014sca}%
  \BibitemOpen
  \bibfield  {author} {\bibinfo {author} {\bibfnamefont {W.-C.}\ \bibnamefont
  {Chen}}\ and\ \bibinfo {author} {\bibfnamefont {J.}~\bibnamefont
  {Piekarewicz}},\ }\href@noop {} {\bibfield  {journal} {\bibinfo  {journal}
  {Phys. Rev.}\ }\textbf {\bibinfo {volume} {C90}},\ \bibinfo {pages} {044305}
  (\bibinfo {year} {2014})}\BibitemShut {NoStop}%
\bibitem [{\citenamefont {Chen}\ and\ \citenamefont
  {Piekarewicz}(2015)}]{Chen:2014mza}%
  \BibitemOpen
  \bibfield  {author} {\bibinfo {author} {\bibfnamefont {W.-C.}\ \bibnamefont
  {Chen}}\ and\ \bibinfo {author} {\bibfnamefont {J.}~\bibnamefont
  {Piekarewicz}},\ }\href@noop {} {\bibfield  {journal} {\bibinfo  {journal}
  {Phys. Lett.}\ }\textbf {\bibinfo {volume} {B748}},\ \bibinfo {pages} {284}
  (\bibinfo {year} {2015})}\BibitemShut {NoStop}%
\bibitem [{\citenamefont {Utama}\ \emph {et~al.}(2016)\citenamefont {Utama},
  \citenamefont {Piekarewicz},\ and\ \citenamefont {Prosper}}]{Utama:2015hva}%
  \BibitemOpen
  \bibfield  {author} {\bibinfo {author} {\bibfnamefont {R.}~\bibnamefont
  {Utama}}, \bibinfo {author} {\bibfnamefont {J.}~\bibnamefont {Piekarewicz}},
  \ and\ \bibinfo {author} {\bibfnamefont {H.~B.}\ \bibnamefont {Prosper}},\
  }\href@noop {} {\bibfield  {journal} {\bibinfo  {journal} {Phys. Rev.}\
  }\textbf {\bibinfo {volume} {C93}},\ \bibinfo {pages} {014311} (\bibinfo
  {year} {2016})}\BibitemShut {NoStop}%
\bibitem [{\citenamefont {Garcia~Ruiz}\ \emph {et~al.}(2016)\citenamefont
  {Garcia~Ruiz} \emph {et~al.}}]{Ruiz:2016gne}%
  \BibitemOpen
  \bibfield  {author} {\bibinfo {author} {\bibfnamefont {R.~F.}\ \bibnamefont
  {Garcia~Ruiz}} \emph {et~al.},\ }\href@noop {} {\bibfield  {journal}
  {\bibinfo  {journal} {Nature Phys.}\ }\textbf {\bibinfo {volume} {12}},\
  \bibinfo {pages} {594} (\bibinfo {year} {2016})}\BibitemShut {NoStop}%
\bibitem [{\citenamefont {Brack}\ and\ \citenamefont
  {Bhaduri}(1997)}]{Brack:1997}%
  \BibitemOpen
  \bibfield  {author} {\bibinfo {author} {\bibfnamefont {M.}~\bibnamefont
  {Brack}}\ and\ \bibinfo {author} {\bibfnamefont {R.~K.}\ \bibnamefont
  {Bhaduri}},\ }\enquote {\bibinfo {title} {Semiclassical physics},}\ \
  (\bibinfo  {publisher} {Addison-Wesley, Reading, MA},\ \bibinfo {year}
  {1997})\BibitemShut {NoStop}%
\bibitem [{\citenamefont {Strutinsky}(1967)}]{Strutinsky:1967}%
  \BibitemOpen
  \bibfield  {author} {\bibinfo {author} {\bibfnamefont {V.~M.}\ \bibnamefont
  {Strutinsky}},\ }\href {\doibase DOI: 10.1016/0375-9474(67)90510-6}
  {\bibfield  {journal} {\bibinfo  {journal} {Nuclear Physics A}\ }\textbf
  {\bibinfo {volume} {95}},\ \bibinfo {pages} {420 } (\bibinfo {year}
  {1967})}\BibitemShut {NoStop}%
\bibitem [{\citenamefont {Piekarewicz}\ \emph {et~al.}(2010)\citenamefont
  {Piekarewicz}, \citenamefont {Centelles}, \citenamefont {Roca-Maza},\ and\
  \citenamefont {Vi\~nas}}]{Piekarewicz:2009av}%
  \BibitemOpen
  \bibfield  {author} {\bibinfo {author} {\bibfnamefont {J.}~\bibnamefont
  {Piekarewicz}}, \bibinfo {author} {\bibfnamefont {M.}~\bibnamefont
  {Centelles}}, \bibinfo {author} {\bibfnamefont {X.}~\bibnamefont
  {Roca-Maza}}, \ and\ \bibinfo {author} {\bibfnamefont {X.}~\bibnamefont
  {Vi\~nas}},\ }\href@noop {} {\bibfield  {journal} {\bibinfo  {journal} {Eur.
  Phys. J.}\ }\textbf {\bibinfo {volume} {A46}},\ \bibinfo {pages} {379}
  (\bibinfo {year} {2010})}\BibitemShut {NoStop}%
\bibitem [{\citenamefont {Garvey}\ and\ \citenamefont
  {Kelson}(1966)}]{Garvey:1966zz}%
  \BibitemOpen
  \bibfield  {author} {\bibinfo {author} {\bibfnamefont {G.~T.}\ \bibnamefont
  {Garvey}}\ and\ \bibinfo {author} {\bibfnamefont {I.}~\bibnamefont
  {Kelson}},\ }\href {\doibase 10.1103/PhysRevLett.16.197} {\bibfield
  {journal} {\bibinfo  {journal} {Phys. Rev. Lett.}\ }\textbf {\bibinfo
  {volume} {16}},\ \bibinfo {pages} {197} (\bibinfo {year} {1966})}\BibitemShut
  {NoStop}%
\bibitem [{\citenamefont {Garvey}\ \emph {et~al.}(1969)\citenamefont {Garvey},
  \citenamefont {Gerace}, \citenamefont {Jaffe}, \citenamefont {Talmi},\ and\
  \citenamefont {Kelson}}]{GARVEY:1969zz}%
  \BibitemOpen
  \bibfield  {author} {\bibinfo {author} {\bibfnamefont {G.~T.}\ \bibnamefont
  {Garvey}}, \bibinfo {author} {\bibfnamefont {W.~J.}\ \bibnamefont {Gerace}},
  \bibinfo {author} {\bibfnamefont {R.~L.}\ \bibnamefont {Jaffe}}, \bibinfo
  {author} {\bibfnamefont {I.}~\bibnamefont {Talmi}}, \ and\ \bibinfo {author}
  {\bibfnamefont {I.}~\bibnamefont {Kelson}},\ }\href {\doibase
  10.1103/RevModPhys.41.S1} {\bibfield  {journal} {\bibinfo  {journal} {Rev.
  Mod. Phys.}\ }\textbf {\bibinfo {volume} {41}},\ \bibinfo {pages} {S1}
  (\bibinfo {year} {1969})}\BibitemShut {NoStop}%
\bibitem [{\citenamefont {Preston}\ and\ \citenamefont
  {Bhaduri}(1993)}]{Preston:1993}%
  \BibitemOpen
  \bibfield  {author} {\bibinfo {author} {\bibfnamefont {M.~A.}\ \bibnamefont
  {Preston}}\ and\ \bibinfo {author} {\bibfnamefont {R.~K.}\ \bibnamefont
  {Bhaduri}},\ }\enquote {\bibinfo {title} {Structure of the nucleus},}\ \
  (\bibinfo  {publisher} {Westview Press},\ \bibinfo {address} {Boulder,
  Colorado},\ \bibinfo {year} {1993})\BibitemShut {NoStop}%
\bibitem [{\citenamefont {Sun}\ \emph {et~al.}(2014)\citenamefont {Sun},
  \citenamefont {Lu}, \citenamefont {Peng}, \citenamefont {Liu},\ and\
  \citenamefont {Zhao}}]{Sun:2014}%
  \BibitemOpen
  \bibfield  {author} {\bibinfo {author} {\bibfnamefont {B.~H.}\ \bibnamefont
  {Sun}}, \bibinfo {author} {\bibfnamefont {Y.}~\bibnamefont {Lu}}, \bibinfo
  {author} {\bibfnamefont {J.~P.}\ \bibnamefont {Peng}}, \bibinfo {author}
  {\bibfnamefont {C.~Y.}\ \bibnamefont {Liu}}, \ and\ \bibinfo {author}
  {\bibfnamefont {Y.~M.}\ \bibnamefont {Zhao}},\ }\href {\doibase
  10.1103/PhysRevC.90.054318} {\bibfield  {journal} {\bibinfo  {journal} {Phys.
  Rev. C}\ }\textbf {\bibinfo {volume} {90}},\ \bibinfo {pages} {054318}
  (\bibinfo {year} {2014})}\BibitemShut {NoStop}%
\bibitem [{\citenamefont {Holt}\ \emph {et~al.}(2012)\citenamefont {Holt},
  \citenamefont {Otsuka}, \citenamefont {Schwenk},\ and\ \citenamefont
  {Suzuki}}]{Holt:2010yb}%
  \BibitemOpen
  \bibfield  {author} {\bibinfo {author} {\bibfnamefont {J.~D.}\ \bibnamefont
  {Holt}}, \bibinfo {author} {\bibfnamefont {T.}~\bibnamefont {Otsuka}},
  \bibinfo {author} {\bibfnamefont {A.}~\bibnamefont {Schwenk}}, \ and\
  \bibinfo {author} {\bibfnamefont {T.}~\bibnamefont {Suzuki}},\ }\href@noop {}
  {\bibfield  {journal} {\bibinfo  {journal} {J. Phys.}\ }\textbf {\bibinfo
  {volume} {G39}},\ \bibinfo {pages} {085111} (\bibinfo {year}
  {2012})}\BibitemShut {NoStop}%
\bibitem [{\citenamefont {Hagen}\ \emph {et~al.}(2012)\citenamefont {Hagen},
  \citenamefont {Hjorth-Jensen}, \citenamefont {Jansen}, \citenamefont
  {Machleidt},\ and\ \citenamefont {Papenbrock}}]{Hagen:2012fb}%
  \BibitemOpen
  \bibfield  {author} {\bibinfo {author} {\bibfnamefont {G.}~\bibnamefont
  {Hagen}}, \bibinfo {author} {\bibfnamefont {M.}~\bibnamefont
  {Hjorth-Jensen}}, \bibinfo {author} {\bibfnamefont {G.~R.}\ \bibnamefont
  {Jansen}}, \bibinfo {author} {\bibfnamefont {R.}~\bibnamefont {Machleidt}}, \
  and\ \bibinfo {author} {\bibfnamefont {T.}~\bibnamefont {Papenbrock}},\
  }\href@noop {} {\bibfield  {journal} {\bibinfo  {journal} {Phys. Rev. Lett.}\
  }\textbf {\bibinfo {volume} {109}},\ \bibinfo {pages} {032502} (\bibinfo
  {year} {2012})}\BibitemShut {NoStop}%
\bibitem [{\citenamefont {Neal}(1996)}]{Neal1996}%
  \BibitemOpen
  \bibfield  {author} {\bibinfo {author} {\bibfnamefont {R.}~\bibnamefont
  {Neal}},\ }\href@noop {} {\emph {\bibinfo {title} {Bayesian Learning of
  Neural Network}}}\ (\bibinfo  {publisher} {Springer},\ \bibinfo {address}
  {New York},\ \bibinfo {year} {1996})\BibitemShut {NoStop}%
\bibitem [{\citenamefont {Bishop}()}]{Bishop1995}%
  \BibitemOpen
  \bibfield  {author} {\bibinfo {author} {\bibfnamefont {C.}~\bibnamefont
  {Bishop}},\ }\href@noop {} {\emph {\bibinfo {title} {Neural Networks for
  Pattern Recognition}}}\ (\bibinfo  {publisher} {Oxford University Press},\
  \bibinfo {address} {Birmingham, UK})\BibitemShut {NoStop}%
\bibitem [{\citenamefont {Haykin}()}]{Haykin1999}%
  \BibitemOpen
  \bibfield  {author} {\bibinfo {author} {\bibfnamefont {S.}~\bibnamefont
  {Haykin}},\ }\href@noop {} {\emph {\bibinfo {title} {Neural Networks: A
  Comprehensive Foundation}}}\ (\bibinfo  {publisher} {Prentice Hall},\
  \bibinfo {address} {Upper Saddle River, NJ})\BibitemShut {NoStop}%
\bibitem [{\citenamefont {Vapnik}()}]{Vapnik1998}%
  \BibitemOpen
  \bibfield  {author} {\bibinfo {author} {\bibfnamefont {V.}~\bibnamefont
  {Vapnik}},\ }\href@noop {} {\emph {\bibinfo {title} {Statistical Learning
  Theory}}}\ (\bibinfo  {publisher} {Wiley-Interscience},\ \bibinfo {address}
  {New York, NY})\BibitemShut {NoStop}%
\bibitem [{\citenamefont {Gazula}\ \emph {et~al.}(1992)\citenamefont {Gazula},
  \citenamefont {Clark},\ and\ \citenamefont {Bohr}}]{Gazula:1992}%
  \BibitemOpen
  \bibfield  {author} {\bibinfo {author} {\bibfnamefont {S.}~\bibnamefont
  {Gazula}}, \bibinfo {author} {\bibfnamefont {J.}~\bibnamefont {Clark}}, \
  and\ \bibinfo {author} {\bibfnamefont {H.}~\bibnamefont {Bohr}},\ }\href
  {\doibase 10.1016/0375-9474(92)90191-L} {\bibfield  {journal} {\bibinfo
  {journal} {Nucl. Phys. A}\ }\textbf {\bibinfo {volume} {540}},\ \bibinfo
  {pages} {1} (\bibinfo {year} {1992})}\BibitemShut {NoStop}%
\bibitem [{\citenamefont {Gernoth}\ \emph {et~al.}(1993)\citenamefont
  {Gernoth}, \citenamefont {Clark}, \citenamefont {Prater},\ and\ \citenamefont
  {Bohr}}]{Gernoth:1993}%
  \BibitemOpen
  \bibfield  {author} {\bibinfo {author} {\bibfnamefont {K.}~\bibnamefont
  {Gernoth}}, \bibinfo {author} {\bibfnamefont {J.}~\bibnamefont {Clark}},
  \bibinfo {author} {\bibfnamefont {J.}~\bibnamefont {Prater}}, \ and\ \bibinfo
  {author} {\bibfnamefont {H.}~\bibnamefont {Bohr}},\ }\href {\doibase
  10.1016/0370-2693(93)90738-4} {\bibfield  {journal} {\bibinfo  {journal}
  {Phys. Lett. B}\ }\textbf {\bibinfo {volume} {300}},\ \bibinfo {pages} {1}
  (\bibinfo {year} {1993})}\BibitemShut {NoStop}%
\bibitem [{\citenamefont {Gernoth}\ and\ \citenamefont
  {Clark}(1995)}]{Gernoth:1995}%
  \BibitemOpen
  \bibfield  {author} {\bibinfo {author} {\bibfnamefont {K.}~\bibnamefont
  {Gernoth}}\ and\ \bibinfo {author} {\bibfnamefont {J.}~\bibnamefont
  {Clark}},\ }\href@noop {} {\bibfield  {journal} {\bibinfo  {journal} {Neural
  Networks}\ }\textbf {\bibinfo {volume} {8}},\ \bibinfo {pages} {291}
  (\bibinfo {year} {1995})}\BibitemShut {NoStop}%
\bibitem [{\citenamefont {Clark}\ \emph {et~al.}(1999)\citenamefont {Clark},
  \citenamefont {Lindenau},\ and\ \citenamefont {Ristig}}]{Clark:1999}%
  \BibitemOpen
  \bibfield  {author} {\bibinfo {author} {\bibfnamefont {J.~W.}\ \bibnamefont
  {Clark}}, \bibinfo {author} {\bibfnamefont {T.}~\bibnamefont {Lindenau}}, \
  and\ \bibinfo {author} {\bibfnamefont {M.}~\bibnamefont {Ristig}},\ }\enquote
  {\bibinfo {title} {Scientific applications of neural nets springer lecture
  notes in physics},}\ \ (\bibinfo  {publisher} {Springer-Verlag},\ \bibinfo
  {address} {Berlin},\ \bibinfo {year} {1999})\ pp.\ \bibinfo {pages}
  {1--96}\BibitemShut {NoStop}%
\bibitem [{\citenamefont {Athanassopoulos}\ \emph {et~al.}(2004)\citenamefont
  {Athanassopoulos}, \citenamefont {Mavrommatis}, \citenamefont {Gernoth},\
  and\ \citenamefont {Clark}}]{Athanassopoulos:2003qe}%
  \BibitemOpen
  \bibfield  {author} {\bibinfo {author} {\bibfnamefont {S.}~\bibnamefont
  {Athanassopoulos}}, \bibinfo {author} {\bibfnamefont {E.}~\bibnamefont
  {Mavrommatis}}, \bibinfo {author} {\bibfnamefont {K.~A.}\ \bibnamefont
  {Gernoth}}, \ and\ \bibinfo {author} {\bibfnamefont {J.~W.}\ \bibnamefont
  {Clark}},\ }\href@noop {} {\bibfield  {journal} {\bibinfo  {journal} {Nucl.
  Phys.}\ }\textbf {\bibinfo {volume} {A743}},\ \bibinfo {pages} {222}
  (\bibinfo {year} {2004})}\BibitemShut {NoStop}%
\bibitem [{\citenamefont {Athanassopoulos}\ \emph {et~al.}(2006)\citenamefont
  {Athanassopoulos}, \citenamefont {Mavrommatis}, \citenamefont {Gernoth},\
  and\ \citenamefont {Clark}}]{Athanassopoulos:2005rc}%
  \BibitemOpen
  \bibfield  {author} {\bibinfo {author} {\bibfnamefont {S.}~\bibnamefont
  {Athanassopoulos}}, \bibinfo {author} {\bibfnamefont {E.}~\bibnamefont
  {Mavrommatis}}, \bibinfo {author} {\bibfnamefont {K.~A.}\ \bibnamefont
  {Gernoth}}, \ and\ \bibinfo {author} {\bibfnamefont {J.~W.}\ \bibnamefont
  {Clark}},\ }\href@noop {} {\emph {\bibinfo {title} {Nuclear mass systematics
  by complementing the finite range droplet model with neural networks}}},\
  edited by\ \bibinfo {editor} {\bibfnamefont {G.}~\bibnamefont {Lalazissis}}\
  and\ \bibinfo {editor} {\bibfnamefont {C.}~\bibnamefont {Moustakidis}}\
  (\bibinfo  {publisher} {Advances in Nuclear Physics, Proceedings of the 15th
  Hellenic Symposium on Nuclear Physics},\ \bibinfo {year} {2006})\ pp.\
  \bibinfo {pages} {65--70}\BibitemShut {NoStop}%
\bibitem [{\citenamefont {Clark}\ and\ \citenamefont
  {Li}(2006)}]{Clark:2006ua}%
  \BibitemOpen
  \bibfield  {author} {\bibinfo {author} {\bibfnamefont {J.~W.}\ \bibnamefont
  {Clark}}\ and\ \bibinfo {author} {\bibfnamefont {H.}~\bibnamefont {Li}},\
  }\href@noop {} {\bibfield  {journal} {\bibinfo  {journal} {Int. J. Mod.
  Phys.}\ }\textbf {\bibinfo {volume} {B20}},\ \bibinfo {pages} {5015}
  (\bibinfo {year} {2006})}\BibitemShut {NoStop}%
\bibitem [{\citenamefont {Costiris}\ \emph {et~al.}(2009)\citenamefont
  {Costiris}, \citenamefont {Mavrommatis}, \citenamefont {Gernoth},\ and\
  \citenamefont {Clark}}]{Costiris:2009}%
  \BibitemOpen
  \bibfield  {author} {\bibinfo {author} {\bibfnamefont {N.~J.}\ \bibnamefont
  {Costiris}}, \bibinfo {author} {\bibfnamefont {E.}~\bibnamefont
  {Mavrommatis}}, \bibinfo {author} {\bibfnamefont {K.~A.}\ \bibnamefont
  {Gernoth}}, \ and\ \bibinfo {author} {\bibfnamefont {J.~W.}\ \bibnamefont
  {Clark}},\ }\href {\doibase 10.1103/PhysRevC.80.044332} {\bibfield  {journal}
  {\bibinfo  {journal} {Phys. Rev. C}\ }\textbf {\bibinfo {volume} {80}},\
  \bibinfo {pages} {044332} (\bibinfo {year} {2009})}\BibitemShut {NoStop}%
\bibitem [{\citenamefont {Bayram}\ \emph {et~al.}(2014)\citenamefont {Bayram},
  \citenamefont {Akkoyun},\ and\ \citenamefont {Kara}}]{Bayram:2013hi}%
  \BibitemOpen
  \bibfield  {author} {\bibinfo {author} {\bibfnamefont {T.}~\bibnamefont
  {Bayram}}, \bibinfo {author} {\bibfnamefont {S.}~\bibnamefont {Akkoyun}}, \
  and\ \bibinfo {author} {\bibfnamefont {S.~O.}\ \bibnamefont {Kara}},\
  }\href@noop {} {\bibfield  {journal} {\bibinfo  {journal} {Annals of Nuclear
  Energy}\ }\textbf {\bibinfo {volume} {63}},\ \bibinfo {pages} {172} (\bibinfo
  {year} {2014})}\BibitemShut {NoStop}%
\bibitem [{\citenamefont {Akkoyun}\ \emph {et~al.}(2013)\citenamefont
  {Akkoyun}, \citenamefont {Bayram}, \citenamefont {Kara},\ and\ \citenamefont
  {Sinan}}]{Akkoyun:2012yf}%
  \BibitemOpen
  \bibfield  {author} {\bibinfo {author} {\bibfnamefont {S.}~\bibnamefont
  {Akkoyun}}, \bibinfo {author} {\bibfnamefont {T.}~\bibnamefont {Bayram}},
  \bibinfo {author} {\bibfnamefont {S.~O.}\ \bibnamefont {Kara}}, \ and\
  \bibinfo {author} {\bibfnamefont {A.}~\bibnamefont {Sinan}},\ }\href@noop {}
  {\bibfield  {journal} {\bibinfo  {journal} {J. Phys.}\ }\textbf {\bibinfo
  {volume} {G40}},\ \bibinfo {pages} {055106} (\bibinfo {year}
  {2013})}\BibitemShut {NoStop}%
\bibitem [{\citenamefont {Angeli}(2004)}]{Angeli:2004}%
  \BibitemOpen
  \bibfield  {author} {\bibinfo {author} {\bibfnamefont {I.}~\bibnamefont
  {Angeli}},\ }\href@noop {} {\bibfield  {journal} {\bibinfo  {journal} {At.
  Data Nucl. Data Tables}\ }\textbf {\bibinfo {volume} {87}},\ \bibinfo {pages}
  {185} (\bibinfo {year} {2004})}\BibitemShut {NoStop}%
\bibitem [{\citenamefont {Angeli}\ and\ \citenamefont
  {Marinova}(2013)}]{Angeli:2013}%
  \BibitemOpen
  \bibfield  {author} {\bibinfo {author} {\bibfnamefont {I.}~\bibnamefont
  {Angeli}}\ and\ \bibinfo {author} {\bibfnamefont {K.}~\bibnamefont
  {Marinova}},\ }\href@noop {} {\bibfield  {journal} {\bibinfo  {journal} {At.
  Data Nucl. Data Tables}\ }\textbf {\bibinfo {volume} {99}},\ \bibinfo {pages}
  {69 } (\bibinfo {year} {2013})}\BibitemShut {NoStop}%
\bibitem [{\citenamefont {Walecka}(1974)}]{Walecka:1974qa}%
  \BibitemOpen
  \bibfield  {author} {\bibinfo {author} {\bibfnamefont {J.~D.}\ \bibnamefont
  {Walecka}},\ }\href@noop {} {\bibfield  {journal} {\bibinfo  {journal}
  {Annals Phys.}\ }\textbf {\bibinfo {volume} {83}},\ \bibinfo {pages} {491}
  (\bibinfo {year} {1974})}\BibitemShut {NoStop}%
\bibitem [{\citenamefont {Boguta}\ and\ \citenamefont
  {Bodmer}(1977)}]{Boguta:1977xi}%
  \BibitemOpen
  \bibfield  {author} {\bibinfo {author} {\bibfnamefont {J.}~\bibnamefont
  {Boguta}}\ and\ \bibinfo {author} {\bibfnamefont {A.~R.}\ \bibnamefont
  {Bodmer}},\ }\href@noop {} {\bibfield  {journal} {\bibinfo  {journal} {Nucl.
  Phys.}\ }\textbf {\bibinfo {volume} {A292}},\ \bibinfo {pages} {413}
  (\bibinfo {year} {1977})}\BibitemShut {NoStop}%
\bibitem [{\citenamefont {Horowitz}\ and\ \citenamefont
  {Serot}(1981)}]{Horowitz:1981xw}%
  \BibitemOpen
  \bibfield  {author} {\bibinfo {author} {\bibfnamefont {C.~J.}\ \bibnamefont
  {Horowitz}}\ and\ \bibinfo {author} {\bibfnamefont {B.~D.}\ \bibnamefont
  {Serot}},\ }\href@noop {} {\bibfield  {journal} {\bibinfo  {journal} {Nucl.
  Phys.}\ }\textbf {\bibinfo {volume} {A368}},\ \bibinfo {pages} {503}
  (\bibinfo {year} {1981})}\BibitemShut {NoStop}%
\bibitem [{\citenamefont {Serot}\ and\ \citenamefont
  {Walecka}(1986)}]{Serot:1984ey}%
  \BibitemOpen
  \bibfield  {author} {\bibinfo {author} {\bibfnamefont {B.~D.}\ \bibnamefont
  {Serot}}\ and\ \bibinfo {author} {\bibfnamefont {J.~D.}\ \bibnamefont
  {Walecka}},\ }\href@noop {} {\bibfield  {journal} {\bibinfo  {journal} {Adv.
  Nucl. Phys.}\ }\textbf {\bibinfo {volume} {16}},\ \bibinfo {pages} {1}
  (\bibinfo {year} {1986})}\BibitemShut {NoStop}%
\bibitem [{\citenamefont {Mueller}\ and\ \citenamefont
  {Serot}(1996)}]{Mueller:1996pm}%
  \BibitemOpen
  \bibfield  {author} {\bibinfo {author} {\bibfnamefont {H.}~\bibnamefont
  {Mueller}}\ and\ \bibinfo {author} {\bibfnamefont {B.~D.}\ \bibnamefont
  {Serot}},\ }\href@noop {} {\bibfield  {journal} {\bibinfo  {journal} {Nucl.
  Phys.}\ }\textbf {\bibinfo {volume} {A606}},\ \bibinfo {pages} {508}
  (\bibinfo {year} {1996})}\BibitemShut {NoStop}%
\bibitem [{\citenamefont {Serot}\ and\ \citenamefont
  {Walecka}(1997)}]{Serot:1997xg}%
  \BibitemOpen
  \bibfield  {author} {\bibinfo {author} {\bibfnamefont {B.~D.}\ \bibnamefont
  {Serot}}\ and\ \bibinfo {author} {\bibfnamefont {J.~D.}\ \bibnamefont
  {Walecka}},\ }\href@noop {} {\bibfield  {journal} {\bibinfo  {journal} {Int.
  J. Mod. Phys.}\ }\textbf {\bibinfo {volume} {E6}},\ \bibinfo {pages} {515}
  (\bibinfo {year} {1997})}\BibitemShut {NoStop}%
\bibitem [{\citenamefont {Lalazissis}\ \emph {et~al.}(1997)\citenamefont
  {Lalazissis}, \citenamefont {Konig},\ and\ \citenamefont
  {Ring}}]{Lalazissis:1996rd}%
  \BibitemOpen
  \bibfield  {author} {\bibinfo {author} {\bibfnamefont {G.~A.}\ \bibnamefont
  {Lalazissis}}, \bibinfo {author} {\bibfnamefont {J.}~\bibnamefont {Konig}}, \
  and\ \bibinfo {author} {\bibfnamefont {P.}~\bibnamefont {Ring}},\ }\href@noop
  {} {\bibfield  {journal} {\bibinfo  {journal} {Phys. Rev.}\ }\textbf
  {\bibinfo {volume} {C55}},\ \bibinfo {pages} {540} (\bibinfo {year}
  {1997})}\BibitemShut {NoStop}%
\bibitem [{\citenamefont {Lalazissis}\ \emph {et~al.}(1999)\citenamefont
  {Lalazissis}, \citenamefont {Raman},\ and\ \citenamefont
  {Ring}}]{Lalazissis:1999}%
  \BibitemOpen
  \bibfield  {author} {\bibinfo {author} {\bibfnamefont {G.~A.}\ \bibnamefont
  {Lalazissis}}, \bibinfo {author} {\bibfnamefont {S.}~\bibnamefont {Raman}}, \
  and\ \bibinfo {author} {\bibfnamefont {P.}~\bibnamefont {Ring}},\ }\href@noop
  {} {\bibfield  {journal} {\bibinfo  {journal} {At. Data Nucl. Data Tables}\
  }\textbf {\bibinfo {volume} {71}},\ \bibinfo {pages} {1} (\bibinfo {year}
  {1999})}\BibitemShut {NoStop}%
\bibitem [{\citenamefont {Horowitz}\ and\ \citenamefont
  {Piekarewicz}(2001)}]{Horowitz:2000xj}%
  \BibitemOpen
  \bibfield  {author} {\bibinfo {author} {\bibfnamefont {C.~J.}\ \bibnamefont
  {Horowitz}}\ and\ \bibinfo {author} {\bibfnamefont {J.}~\bibnamefont
  {Piekarewicz}},\ }\href@noop {} {\bibfield  {journal} {\bibinfo  {journal}
  {Phys. Rev. Lett.}\ }\textbf {\bibinfo {volume} {86}},\ \bibinfo {pages}
  {5647} (\bibinfo {year} {2001})}\BibitemShut {NoStop}%
\bibitem [{\citenamefont {Todd-Rutel}\ and\ \citenamefont
  {Piekarewicz}(2005)}]{Todd-Rutel:2005fa}%
  \BibitemOpen
  \bibfield  {author} {\bibinfo {author} {\bibfnamefont {B.~G.}\ \bibnamefont
  {Todd-Rutel}}\ and\ \bibinfo {author} {\bibfnamefont {J.}~\bibnamefont
  {Piekarewicz}},\ }\href@noop {} {\bibfield  {journal} {\bibinfo  {journal}
  {Phys. Rev. Lett}\ }\textbf {\bibinfo {volume} {95}},\ \bibinfo {pages}
  {122501} (\bibinfo {year} {2005})}\BibitemShut {NoStop}%
\bibitem [{\citenamefont {Mumpower}\ \emph {et~al.}(2016)\citenamefont
  {Mumpower}, \citenamefont {Surman}, \citenamefont {McLaughlin},\ and\
  \citenamefont {Aprahamian}}]{Mumpower:2015ova}%
  \BibitemOpen
  \bibfield  {author} {\bibinfo {author} {\bibfnamefont {M.~R.}\ \bibnamefont
  {Mumpower}}, \bibinfo {author} {\bibfnamefont {R.}~\bibnamefont {Surman}},
  \bibinfo {author} {\bibfnamefont {G.~C.}\ \bibnamefont {McLaughlin}}, \ and\
  \bibinfo {author} {\bibfnamefont {A.}~\bibnamefont {Aprahamian}},\
  }\href@noop {} {\bibfield  {journal} {\bibinfo  {journal} {Prog. Part. Nucl.
  Phys.}\ }\textbf {\bibinfo {volume} {86}},\ \bibinfo {pages} {86} (\bibinfo
  {year} {2016})}\BibitemShut {NoStop}%
\bibitem [{\citenamefont {Aberg}(2002)}]{Aberg:2012}%
  \BibitemOpen
  \bibfield  {author} {\bibinfo {author} {\bibfnamefont {S.}~\bibnamefont
  {Aberg}},\ }\href@noop {} {\bibfield  {journal} {\bibinfo  {journal}
  {Nature}\ }\textbf {\bibinfo {volume} {417}},\ \bibinfo {pages} {499}
  (\bibinfo {year} {2002})}\BibitemShut {NoStop}%
\bibitem [{\citenamefont {Barea}\ \emph {et~al.}(2005)\citenamefont {Barea},
  \citenamefont {Frank}, \citenamefont {Hirsch},\ and\ \citenamefont
  {Van~Isacker}}]{Barea:2005fz}%
  \BibitemOpen
  \bibfield  {author} {\bibinfo {author} {\bibfnamefont {J.}~\bibnamefont
  {Barea}}, \bibinfo {author} {\bibfnamefont {A.}~\bibnamefont {Frank}},
  \bibinfo {author} {\bibfnamefont {J.~G.}\ \bibnamefont {Hirsch}}, \ and\
  \bibinfo {author} {\bibfnamefont {P.}~\bibnamefont {Van~Isacker}},\
  }\href@noop {} {\bibfield  {journal} {\bibinfo  {journal} {Phys. Rev. Lett.}\
  }\textbf {\bibinfo {volume} {94}},\ \bibinfo {pages} {102501} (\bibinfo
  {year} {2005})}\BibitemShut {NoStop}%
\bibitem [{\citenamefont {Hornik}\ \emph {et~al.}(1989)\citenamefont {Hornik},
  \citenamefont {Stinchcombe},\ and\ \citenamefont {White}}]{Hornik:1989}%
  \BibitemOpen
  \bibfield  {author} {\bibinfo {author} {\bibfnamefont {K.}~\bibnamefont
  {Hornik}}, \bibinfo {author} {\bibfnamefont {M.}~\bibnamefont {Stinchcombe}},
  \ and\ \bibinfo {author} {\bibfnamefont {H.}~\bibnamefont {White}},\
  }\href@noop {} {\bibfield  {journal} {\bibinfo  {journal} {Neural Networks}\
  }\textbf {\bibinfo {volume} {2}},\ \bibinfo {pages} {359 } (\bibinfo {year}
  {1989})}\BibitemShut {NoStop}%
\bibitem [{\citenamefont {Titterington}(2004)}]{Titterington:2004}%
  \BibitemOpen
  \bibfield  {author} {\bibinfo {author} {\bibfnamefont {D.~M.}\ \bibnamefont
  {Titterington}},\ }\href@noop {} {\bibfield  {journal} {\bibinfo  {journal}
  {Statist. Sci.}\ }\textbf {\bibinfo {volume} {19}},\ \bibinfo {pages} {128}
  (\bibinfo {year} {2004})}\BibitemShut {NoStop}%
\bibitem [{\citenamefont {Stone}(2013)}]{Stone:2013}%
  \BibitemOpen
  \bibfield  {author} {\bibinfo {author} {\bibfnamefont {J.~V.}\ \bibnamefont
  {Stone}},\ }\enquote {\bibinfo {title} {Bayes' rule: A tutorial introduction
  to bayesian analysis},}\ \ (\bibinfo  {publisher} {Sebtel Press},\ \bibinfo
  {address} {Sheffield, UK},\ \bibinfo {year} {2013})\ \bibinfo {edition}
  {1st}\ ed.\BibitemShut {Stop}%
\bibitem [{\citenamefont {MacKay}(1995)}]{Mackay:1995}%
  \BibitemOpen
  \bibfield  {author} {\bibinfo {author} {\bibfnamefont {D.~J.}\ \bibnamefont
  {MacKay}},\ }\href@noop {} {\bibfield  {journal} {\bibinfo  {journal}
  {Nuclear Instruments and Methods in Physics Research Section A}\ }\textbf
  {\bibinfo {volume} {354}},\ \bibinfo {pages} {73 } (\bibinfo {year}
  {1995})}\BibitemShut {NoStop}%
\bibitem [{\citenamefont {MacKay}(1999)}]{Mackay:1999}%
  \BibitemOpen
  \bibfield  {author} {\bibinfo {author} {\bibfnamefont {D.~J.}\ \bibnamefont
  {MacKay}},\ }\href@noop {} {\bibfield  {journal} {\bibinfo  {journal} {Neural
  Computation}\ }\textbf {\bibinfo {volume} {11}},\ \bibinfo {pages} {1035}
  (\bibinfo {year} {1999})}\BibitemShut {NoStop}%
\bibitem [{\citenamefont {Richter}\ and\ \citenamefont
  {Brown}(2003)}]{Richter:2003wi}%
  \BibitemOpen
  \bibfield  {author} {\bibinfo {author} {\bibfnamefont {W.~A.}\ \bibnamefont
  {Richter}}\ and\ \bibinfo {author} {\bibfnamefont {B.~A.}\ \bibnamefont
  {Brown}},\ }\href {\doibase 10.1103/PhysRevC.67.034317} {\bibfield  {journal}
  {\bibinfo  {journal} {Phys. Rev.}\ }\textbf {\bibinfo {volume} {C67}},\
  \bibinfo {pages} {034317} (\bibinfo {year} {2003})}\BibitemShut {NoStop}%
\bibitem [{\citenamefont {Wang}\ \emph {et~al.}(2012)\citenamefont {Wang},
  \citenamefont {Audi}, \citenamefont {Wapstra}, \citenamefont {Kondev},
  \citenamefont {MacCormick}, \citenamefont {Xu},\ and\ \citenamefont
  {Pfeiffer}}]{AME:2012}%
  \BibitemOpen
  \bibfield  {author} {\bibinfo {author} {\bibfnamefont {M.}~\bibnamefont
  {Wang}}, \bibinfo {author} {\bibfnamefont {G.}~\bibnamefont {Audi}}, \bibinfo
  {author} {\bibfnamefont {A.}~\bibnamefont {Wapstra}}, \bibinfo {author}
  {\bibfnamefont {F.}~\bibnamefont {Kondev}}, \bibinfo {author} {\bibfnamefont
  {M.}~\bibnamefont {MacCormick}}, \bibinfo {author} {\bibfnamefont
  {X.}~\bibnamefont {Xu}}, \ and\ \bibinfo {author} {\bibfnamefont
  {B.}~\bibnamefont {Pfeiffer}},\ }\href {\doibase 10.1088/1674-1137/36/12/003}
  {\bibfield  {journal} {\bibinfo  {journal} {Chinese Phys. C}\ }\textbf
  {\bibinfo {volume} {36}},\ \bibinfo {pages} {1603} (\bibinfo {year}
  {2012})}\BibitemShut {NoStop}%
\bibitem [{\citenamefont {Caurier}\ \emph {et~al.}(2001)\citenamefont
  {Caurier}, \citenamefont {Langanke}, \citenamefont {Martinez-Pinedo},
  \citenamefont {Nowacki},\ and\ \citenamefont {Vogel}}]{Caurier:2001np}%
  \BibitemOpen
  \bibfield  {author} {\bibinfo {author} {\bibfnamefont {E.}~\bibnamefont
  {Caurier}}, \bibinfo {author} {\bibfnamefont {K.}~\bibnamefont {Langanke}},
  \bibinfo {author} {\bibfnamefont {G.}~\bibnamefont {Martinez-Pinedo}},
  \bibinfo {author} {\bibfnamefont {F.}~\bibnamefont {Nowacki}}, \ and\
  \bibinfo {author} {\bibfnamefont {P.}~\bibnamefont {Vogel}},\ }\href@noop {}
  {\bibfield  {journal} {\bibinfo  {journal} {Phys. Lett.}\ }\textbf {\bibinfo
  {volume} {B522}},\ \bibinfo {pages} {240} (\bibinfo {year}
  {2001})}\BibitemShut {NoStop}%
\bibitem [{\citenamefont {Saperstein}\ and\ \citenamefont
  {Tolokonnikov}(2011)}]{Saperstein:2011}%
  \BibitemOpen
  \bibfield  {author} {\bibinfo {author} {\bibfnamefont {E.~E.}\ \bibnamefont
  {Saperstein}}\ and\ \bibinfo {author} {\bibfnamefont {S.~V.}\ \bibnamefont
  {Tolokonnikov}},\ }\href@noop {} {\bibfield  {journal} {\bibinfo  {journal}
  {Physics of Atomic Nuclei}\ }\textbf {\bibinfo {volume} {74}},\ \bibinfo
  {pages} {1277} (\bibinfo {year} {2011})}\BibitemShut {NoStop}%
\end{thebibliography}%

\end{document}